\documentclass[fleqn,usenatbib]{mnras}
\usepackage{newtxtext,newtxmath,anyfontsize}
\usepackage[T1]{fontenc}

\newcommand{\rev}{}
\newcommand{\revv}{}
\newcommand{\revvv}{}

\DeclareRobustCommand{\VAN}[3]{#2}
\let\VANthebibliography\thebibliography
\def\thebibliography{\DeclareRobustCommand{\VAN}[3]{##3}\VANthebibliography}
\usepackage{graphicx}	% Including figure files
\usepackage{amsmath}	% Advanced maths commands
\title[Sublimative rotational breakup around WDs]{Producing planetary debris exterior to white dwarf Roche radii through sublimative rotational fission}

\author[]{Dimitri Veras$^{1,2,3}$\thanks{E-mail: dimitri.veras@aya.yale.edu},
Jordan K. Steckloff$^{4,5}$,
Kathryn Volk$^{4}$
\\
$^{1}$Centre for Exoplanets and Habitability, University of Warwick, Coventry CV4 7AL, UK
\\
$^{2}$Centre for Space Domain Awareness, University of Warwick, Coventry CV4 7AL, UK
\\
$^{3}$Department of Physics, University of Warwick, Coventry CV4 7AL, UK
\\
$^{4}$The Planetary Science Institute, Tucson, AZ, USA
\\
$^{5}$Department of Aerospace Engineering and Engineering Mechanics, University of Texas at Austin, Austin, TX, USA
}

\pubyear{\the\year{}}

\begin{document}
\label{firstpage}
\pagerange{\pageref{firstpage}--\pageref{lastpage}}
\maketitle

% Abstract of the paper
\begin{abstract}
The majority of white dwarfs that host periodic transiting planetary debris do so at distances that exceed the rubble-pile Roche limit, in disagreement with canonical formation models that focus on the tidal disruption of minor planets. Here, we quantify the conditions by which rotational fission due to sublimative outgassing {\revvv (“SYORP” break-up)} can occur outside of the Roche sphere {\revvv in the distance range of 1-5 Roche radii}. We use the Many Materials Orbital Sublimation (MaMOS) model to quantify the outgassing properties of three representative types of planetary materials: cores (iron), mantles (forsterite olivine) and comets (water ice), and characterise the resulting spin-up rate analytically {\revvv by adopting SYORP coefficients in the range of $10^{-5}-10^{-3}$}. We then compare this rate to that generated by the radiative YORP effect {\revvv with YORP coefficients of $10^{-3}-10^{-2}$, and focus on planetesimals with radii of 0.1, 1.0, and 10 km}. We find that for white dwarf cooling ages of up to $\sim 1$~Gyr, sublimative fission of planetesimals and fragments $\lesssim 0.1$~km in size due to water ice outgassing occur on observable timescales (within 10~yr),  regardless if the spin-up is monotonic or stochastic. Further, these timescales are orders of magnitude shorter than the corresponding YORP fission timescales. For drier planetesimals, both iron and forsterite outgassing can be effective at 10-100~Myr cooling ages. {\revvv Our results do not substantially differ for strengthless rubble piles versus objects with 1~kPa of internal strength.} These findings add to growing evidence that gravitationally scattered comets and asteroids do not need to adopt pericentres within a white dwarf's Roche radius to eventually enrich, or pollute, the star with metals. 
\end{abstract}


%%%%%%%%%%%%%%%%%%%%%%%%%%%%%%%%%%%%%%%%%%%%%%%%%%
%%%%%%%%%%%%%%%%%%%%%%%%%%%%%%%%%%%%%%%%%%%%%%%%%%

%%%%%%%%%%%%%%%%% BODY OF PAPER %%%%%%%%%%%%%%%%%%

\section{Introduction}

Current observations of white dwarf planetary systems can be split into three major categories: (i) the planets themselves \citep{thoetal1993,sigetal2003,luhetal2011,ganetal2019,vanderburgetal2020,blaetal2021,limetal2024,zhaetal2024}, (ii) the ubiquitous metal enrichment (or ``pollution") seen in the white dwarf atmospheres \citep[e.g.][]{wiletal2024,xuetal2024,bonsor2026}, and (iii) discs of debris orbiting the white dwarfs well outside of these atmospheres but well within the Mercury-Sun distance. 

These debris discs represent the transition stage between the other two categories \citep{malamud2026}; they form from the remnants of the initial planetary system and later accrete onto the white dwarf. Hence, the formation, evolution and destruction of the discs inform the structure and dynamics of the initial planetary system, as well as the composition and frequency of the metal enrichment in the white dwarf atmospheres.

From 1987-2015, white dwarf debris discs were discovered primarily through two observational features: infrared excesses on spectral energy distributions \citep[e.g.][]{zucbec1987,graetal1990,faretal2009,faretal2010,xuetal2020,oweetal2023,wanetal2023,wanetal2026,faretal2025,nooetal2025} and gaseous emission lines \citep[e.g.][]{ganetal2006,ganetal2007,denetal2020,manetal2020,meletal2020,genetal2021,rogetal2025}. However, in 2015, a step change occurred with the first detection of large ($\sim$100 km) transiting planetary debris \citep{vanetal2015,vanrap2018}.

This debris was discovered around white dwarf WD~1145+017, which also showcased infrared excess and gaseous spectral signatures. Further, the atmosphere of WD~1145+017 was shown to be enriched with metals. Hence, for the first time, the metals, the debris, the dust and the gas were all linked within the same system. Additionally, the orbital period of the debris was about 4.5~hr, corresponding to the location of the rigid, spinning rubble-pile Roche limit of the white dwarf, reinforcing a long-assumed theory that rubble-pile-like asteroids commonly break up at this radial location \citep{jura2003}.

Since 2015, observations of five more white dwarfs with transiting debris have established well-measured orbital periodicities \citep{vanderboschetal2020,vanderboschetal2021,faretal2022,guietal2025,heretal2025} \footnote{Further, there are many more known white dwarfs with transiting debris whose periodicities have not yet been established \citep{guietal2021,bhaetal2025}.}. However, the distribution of these orbital periods has been surprising. Instead of being clustered around 4.5 hours, the periods range from 5-25 hours, and in one case, feature a period of 107 days \citep{vanderboschetal2020}. Even the 5-25 hour period range corresponds to semi-major axes that vary significantly ($\approx$1-4 Roche radii), given that the masses of the white dwarfs in all cases are confined to within the range $0.5-0.8M_{\odot}$, with respective Roche limits that vary by only $\sim20$ per$\,$cent under otherwise identical situations.

Agreed-upon explanations for the dynamical origin of {\revv these unexpectedly long periodicities remain outstanding}, although several ideas have been proposed\footnote{{\revv Producing debris exterior to the Roche radius can be accomplished with sufficiently massive objects which exceed about one lunar mass, with $\gtrsim 10^{23}$~kg, that intersect the white dwarf Roche sphere \citep{malper2020a,malper2020b,toretal2026}. However, the resulting debris itself does not translate into periodicities unless clumping or second-generation formation occurs; in the later case, the periodicites would extend out to only just beyond 2$R_{\rm Roche}$ \citep{vanlieshoutetal2018}. Second-generation formation in this manner is assumed to be uncommon because such high impactor masses are probably not common enough to explain the metal-enrichment in $\sim$40-50 per\,cent of all white dwarfs \citep{koeetal2014,ouletal2024}, and the mass of the parent body in the WD~1145+017 system has been inferred to be three orders-of-magnitude lower \citep{rapetal2016,guretal2017}.}}. One set of potential mechanisms invoke ejecta from a minor planet close to, {\revv but not at, the Roche radius of the} white dwarf. The ejecta could arise from:

\begin{enumerate}

\item impacts from external bombardments \citep{verkur2020}, 

\item sesquinary or returning impacts \citep{vercuk2025}, 

\item or volcanism on the minor planet itself \citep{kisetal2023,lietal2025b}. 

\end{enumerate}

Another set of mechanisms relies on the breakup, or fission, of the minor planet through some internal process. Four flavours of this mechanism include:

\begin{enumerate}

\item Fission due to spin and orbital angular momentum exchange. \cite{vermcdmak2020} illustrated that for minor planets which are very physically elongated (i.e. where one dimension is five times the length of another), the coupling between their orbital and spin angular momentum may break them apart through rotational fission when they reside up to three times distant from the rubble-pile Roche limit. However, for minor planets where the length of all three of their axes are within a factor of two of one another\footnote{{\rev The elongation of minor planets orbiting white dwarfs is unconstrained. Arguably the only intact exo-asteroid which has been detected \citep{manetal2019} was found indirectly, without any shape constraints, and is anyway more likely to represent a high-density, high-strength planetary core fragment. Theoretically, giant branch evolution may have elongated asteroids which survived main-sequence evolution \citep{katz2018}.}}, this mechanism would struggle to break them apart beyond twice the rubble-pile Roche limit.

\item Fission due to radiative torques. This mechanism is well-known in solar system science as the Yarkovsky-O'Keefe-Radzievskii-Paddack effect, more commonly known as the YORP effect \citep[e.g.][]{rubincam2000,botetal2006,voketal2015,jewitt2025}. It is most effective in cases of hotter white dwarfs and when the minor planet is both already spinning quickly and is of a sufficiently small size ($<100$~km) for the effect to take hold \citep{verbirzam2022}.

\item Fission through thermal cracking. This process is also temperature-dependent, and has recently been explored for debris orbiting white dwarfs \citep{sheser2023}. In this case, thermal stresses can begin to tear apart the surface layers of boulders of size $10^{-2}-10^1$~m, and do so at distances of up to tens to hundreds of Roche radii. Indeed, thermal fatigue and fracturing have been proposed to drive the observed ejection of small $\sim$cm-sized chunks from asteroid 101955 Bennu, the target of the OSIRIS-ReX mission \citep{lauetal2019,moletal2020}.

\item Fission due to sublimative torques. To our knowledge, this mechanism has not yet been explored in detail, and is the topic of this work. In planetary systems around white dwarfs, sublimation has already been investigated in multiple other contexts \citep[e.g.][]{jurxu2010,juretal2012,vereggganSUB2015,malper2016,malper2017a,malper2017b,katz2018,steckloffetal2021a,okuetal2023}, none of which specifically focus on how outgassing changes the rotation rate of minor planets. Like with the YORP effect, sublimative fission will likely depend on the luminosity, and hence the cooling age, of the white dwarf. Hence, understanding how these effects compare is important, especially as both could act on similarly-sized minor planets.

\end{enumerate}

To explore sublimative fission, we use the Many Materials Orbital Sublimation (MaMOS) model, which is described in Section \ref{sec:MaMOS}. We then apply the model in Section \ref{sec:results}, where we present our results; these are discussed in Section \ref{sec:discussion}. The paper concludes with a summary in Section \ref{sec:summary}.

Throughout the paper, we describe the subliming body as the ``small object" or ``small body". In this way, we can keep the treatment general and analyse different material applications. 

\section{MaMOS spin-up model}\label{sec:MaMOS}

\cite{stejac2016}, and later \cite{safetal2021} and \cite{steckloffetal2021b}, developed the Many Materials Orbital Sublimation (MaMOS) model from the YORP-based formalism of \cite{scheeres2007} to understand sublimative torques as the sublimative analogue to the YORP effect. They named the effect the Sublimation-YORP, or SYORP, effect. 

\subsection{Equation for spin-up}

The SYORP formalism describes the angular acceleration $d\omega/dt$ experienced by a homogeneous, predominantly spherical object as 
\begin{equation}
     \left(\frac{d\omega}{dt}\right)_{\rm sub}
=
\frac{3P_S C_S}{4\pi \rho R^2},
\label{Eq:spinsub}
\end{equation}
\noindent{}where $\rho$ and $R$ are the density and radius of the small body. $C_S$ is the dimensionless so-called SYORP coefficient -- usually treated as a constant -- which describes the fraction of sublimative momentum flux that is directed tangential to the surface of the small body and thus exerts a torque on that body \citep{stejac2016}. $P_S$ is the dynamic sublimation back reaction pressure on the small body, and is the main source of complexity in the model. This quantity describes the pressure that a sublimating object exerts due to gas escaping to vacuum \citep{steetal2015}. 

The value of $P_{\rm S}$ is computed by solving the following system of equations, which describe the energy balance at the surface of a subliming object. The equations incorporate the energy flux from the substellar surface from stellar heating ($I_{\rm stellar}$), radiative cooling ($I_{\rm radiative}$), and sublimative cooling ($I_{\rm sublimative}$), such that
\begin{equation}
    I_{\rm stellar}=I_{\rm sublimative}+I_{\rm radiative},
    \label{Eq:MAMOS1}
    \end{equation}
\noindent{}where
    \begin{equation}
    I_{\rm stellar}=(1-A)\frac{L_{\rm WD}}{4\pi r^2},
\end{equation}
\begin{equation}
    I_{\rm radiative} = \epsilon\varsigma_{SB}T^4,
\end{equation}
\noindent{}and
\begin{equation}
    I_{\rm sublimative}=\lambda \frac{dm}{dt}=\lambda\alpha(T)\sqrt{\frac{m_{\rm mol}}{2\pi \mathcal{R}T}}P_{\rm ref}e^{\frac{\lambda}{\mathcal{R}}(\frac{1}{T_{\rm ref}}-\frac{1}{T})}.
    \label{Eq:Isub}
\end{equation}

In these equations, $A$ is the surface albedo of the small body, $L_{\rm WD}$ is the luminosity of the white dwarf, $r$ is the astrocentric distance of the small body from the star, $\epsilon$ is the emissivity of the surface of the small body, $\varsigma_{SB}$ is the Stefan-Boltzmann constant, $T$ is the surface temperature of the small body, $\lambda$ is the material-dependent latent heat of sublimation (in energy/mass), $m$ is the mass of the small body, $\alpha(T)$ is the temperature-dependent sublimation coefficient of the material and is typically close to unity \citep{langmuir1913}, $m_{\rm mol}$ is the molar mass of the subliming material, $\mathcal{R}$ is the ideal gas constant, and $P_{\rm ref}$ and $T_{\rm ref}$ are a pair of experimentally determined reference vapor pressure and temperature \citep{steetal2015}. 

Our numerical model solves this system of equations (\ref{Eq:MAMOS1}-\ref{Eq:Isub}) and outputs a lookup table with several quantities. These include the temperature, sublimative mass flux ($dm/dt$) and the dynamic sublimation pressure via
\begin{equation}
P_S = v_{\rm therm}\frac{dm}{dt},
\label{Eq:Ps}
\end{equation}
where $v_{\rm therm}$ is the mean of the magnitude of the thermal speeds of the escaping molecules
\begin{equation}
    v_{\rm therm}=\sqrt{\frac{8\mathcal{R}T}{\pi m_{\rm mol}}}.
    \label{Eq:vtherm}
\end{equation}

Equation (\ref{Eq:spinsub}) can be re-written, though equations (\ref{Eq:Isub}), (\ref{Eq:Ps}) and (\ref{Eq:vtherm}), as
\begin{equation}
\left(\frac{d\omega}{dt} \right)_{\rm sub} =\frac{3C_{\rm S}\alpha P_{\rm ref}}
{2 \pi^2 \rho R^2}
e^{\frac{\lambda}{\mathcal{R}}(\frac{1}{T_{\rm ref}}-\frac{1}{T})}.
\label{Eq:SubSpin}
\end{equation}
\noindent{}Equation (\ref{Eq:SubSpin}) is the key equation that we use to compute sublimative spin-up. 

\subsection{Scaling with $C_{\rm S}$}
\label{CSScale}

{\rev The physical reason why $C_S$ is nonzero is because outgassing is at least slightly anisotropic. Although the overall sublimation pattern from comets is radially symmetric, it is the minor, tangential components (representing less than 0.1 per cent of the momentum flux) that generate a net torque that spins up the object. Both isotropic and anisotropic sublimation occur concurrently. However, we only concern ourselves with the tangential component, as that is the one that drives spin-up and generates a non-zero value of $C_S$.}

Despite sublimating objects such as comets exhibiting wildly different sublimative activity levels that vary by orders of magnitude, observed objects nevertheless appear to spin up by a relatively constant amount due to randomly oriented sources of torque, tending to cancel one another out as activity levels increase \citep{sammue2013,muesam2018,knikok2024}. This constancy suggests that {\it all} sublimating small bodies will have values for the SYORP coefficient $C_S$ that are similar to one another, as confimed by \cite{muesam2018}, \cite{stesam2018}, and \cite{steckloffetal2021b}. In particular, \cite{steckloffetal2021b}  found that the SYORP coefficients of measured comets vary between $10^{-5}-10^{-4}$, although particularly fresh, active, and/or angular small bodies may plausibly have  $C_S$ values of up to $\sim10^{-3}$ \citep{stejac2016,safetal2021}, and are therefore worth considering as well.

{\rev Having established the allowable range for $C_{\rm S}$ ($10^{-5}-10^{-3}$), we now consider how this parameter could be a function of time. Determination of this time dependence for a particular object can occur only if (i) we have detailed data for that object in terms of its mineralogical and topological distributions, or if (ii) we have observed the mass loss from the object for long enough and in sufficient detail to obtain an empirical estimate. Hence, in white dwarf planetary systems, we currently cannot determine the time evolution of $C_{\rm S}$.}

{\rev Instead, we may speculate on the time evolution of $C_{\rm S}$ by considering solar system examples. Like SYORP, the YORP effect is often modelled with a spin-up timescale that is proportional to a constant, $\chi$, which is determined by the detailed physical characteristics of the asteroid. Over the Myr timescales often considered, $\chi$ has been shown to vary with time. In fact, $\chi$ better reproduces observational results when treated as a stochastic function of time, with rotational fission timescales being about 10 times longer than in the constant $\chi$ case \citep{beretal2026}.}

{\rev If $C_{\rm S}$ also is a function of time, then the variation would be restricted to the range $10^{-5}-10^{-3}$. Hence, we can determine the range of possible outcomes by adopting $C_{\rm S} = 10^{-4}$ for our simulations and then simply increasing or decreasing the resulting timescale by a factor of 10. %We refer to the simulation outcomes for $C_{\rm S} = 10^{-4}$ as "monotonic spin-up", and label outcomes for $C_{\rm S} = 10^{-4}$ 
}

\subsection{Scaling with other variables}

Here, we assume circular orbits for this initial exploration\footnote{To analyse eccentric orbits, one would need to consider the values in equation (\ref{Eq:SubSpin}) that are dependent on separation ($T$ and $\alpha(T)$). The value of $C_{\rm S}$ could also vary with separation, but would be restricted in the $10^{-5}-10^{-3}$ range.}. Doing so represents a reasonable starting point {\revv from both theoretical and observational points-of-view. Objects that are 10-100~km in size could retain volatiles from formation and throughout the giant branch phases of stellar evolution \citep{malper2016,malper2017a,malper2017b}. Subsequently, after being scattered close to the white dwarf, these objects will harbour orbits that shrink and circularise in the distance regime studied here, within a few white dwarf Roche radii \citep{lietal2025a}, where the objects can fragment \citep{mcdver2021} and expose volatiles to the surface. Observationally,} the majority of white dwarfs with transiting debris indicate that the debris is on circular or near-circular orbits, constraining the small body to have broken up, at least around WD~1145+017, in a similarly circular orbit  \citep{vercarleietal2017,duvetal2020}. 

Given the assumption of circularity, we importantly note that the right-hand-side of the equation is effectively constant at the level of approximation needed here. {\rev Additionally,} as bits of the small object are outgassed, both $C_{\rm S}$ and $R$ will change, but negligibly over secular timescales. Observations of comets have found that they typically lose $\sim$1 m of material from active surfaces over an orbit \citep{grolam2003,keletal2015}, a very small layer compared to the $\sim$1 km size of the nucleus. 
%In any case, our methodology will err toward longer timescales due to this approximation, and thus represents a conservative estimate. 

Further, the value of $T$ will change as the white dwarf cools. However, the white dwarf cooling timescale (Myr-to-Gyr) is many orders of magnitude greater than {\revv both} the orbital period of debris just outside of the Roche sphere (tens of hours), {\revv and the disruption timescales that we will later calculate}. Hence, $T$ may also be treated as constant unless the small body orbit is at least moderately eccentric. Even in the case of highly eccentric orbits, these results nevertheless bound the spin up timescales. {\rev Because each simulation assumes a fixed white dwarf cooling age and a circular orbit, we treat $T$ as constant within each simulation.}

In equation (\ref{Eq:SubSpin}), the only variable which represents an output from the MaMOS model is $T$. All other variables either represent constants ($\mathcal{R}$) or input model parameters ($R, C_{\rm S}, \alpha, \lambda, P_{\rm ref}, T_{\rm ref}, \rho$). Of this last set, the final five variables are material-dependent. 

\subsection{Materials adopted}

In this work, we consider three different types of materials: iron, water ice, and forsterite. Our choice is motivated by (i) the availability of numerical parameters for these materials, and (ii) observational evidence for all three materials in white dwarf planetary systems \citep[e.g.][]{faretal2013,radetal2015,xuetal2017,holetal2018,hosetal2020,putxu2021,johetal2022,wiletal2025,sahetal2025}. These three materials can be thought of as proxies for, respectively, core material, comets, and mantle material.

We display the parameters for each of these materials in Table \ref{TableInParam_models}. For those parameters with ranges, we adopt the following values in our subsequent computations: $\alpha_{\rm Forsterite}=0.10$, $\alpha_{\rm Water \ Ice}=0.50$.

\begin{table*}
\caption {\label{TableInParam_models} Physical properties of the materials considered in this work. In order to obtain realistic representations of these materials and their host bodies, we constructed characteristic exo-planetesimals based on analogous solar system objects. We utilised the sample from the lunar exploration mission, Change'e 6, as most representative of the Lunar mantle due to provenance from the Moon's South Pole-Aitken Basin.}
\begin{tabular}{cccccccc}
 \vspace{-0.1cm} \\
  Representative object & Material & $P_{\rm Ref}$ & $T_{\rm Ref}$ & $\lambda$ & $\alpha$ & $
 C_{\rm s}$  & $\rho$ 
\\ 
  & & (Pa) & (K) & (J/kg) &  &  & (kg/m$^3$) 
\\
 \cline{1-8}  
\\
 Iron meteorite (core material) & Iron & 14.52 & 1867 & 340800 & 1 & $10^{-5}-10^{-3}$  & 7874   
\\
 \vspace{-0.2cm} \\
 Lunar mantle (Change'e 6) & Forsterite & 0.421 & 1998 & 547000 & 0.04 - 0.19 \citep{steckloffetal2021a} &  $10^{-5}-10^{-3}$ & 3000    \\
 \vspace{-0.2cm} \\
 Comet & Water ice & 0.01973 & 187.02 & 56200 & ~0.14 - 1 \citep{gunetal2011} &  $10^{-5}-10^{-3}$  & 500  \\
\end{tabular}
\begin{tabbing}
\end{tabbing}
\end{table*}

\section{Results}\label{sec:results}

We now use the MaMOS spin-up model to illustrate which parameter choices allow for sublimative breakup outside of the white dwarf Roche sphere, while comparing with YORP-based fission. First we need to establish the threshold for breakup (Section \ref{sec:breakup}). Then, we need to quantify YORP spin-up (Section \ref{sec:YORP}), and relate the age and luminosity of the white dwarf to the MaMOS and YORP models (Section \ref{sec:WDAge}) before actually applying the models (Section \ref{sec:apply}).

\subsection{Critical breakup speed}\label{sec:breakup}

When the value of $\omega(t)$ exceeds a critical value $\omega_{\rm crit}$, then the object will break apart. One formulation for the breakup spin is \citep{sansch2014,scheeres2018} 
\begin{equation}
\omega_{\rm crit} = \sqrt{\frac{Gm}{R^3} + \frac{16\pi\sigma R}{9m}}
=\sqrt{\frac{4\pi \rho G}{3} + \frac{4\sigma }{3 \rho R^2}}, 
\label{Eq:critDimitri}
\end{equation}
\noindent{}where $\sigma$ is the object's tensile strength. A different formulation which takes into account the tri-axial nature of many minor planets can be found in equations 6 and A1-A13 of \cite{vermcdmak2020}. Regardless, equation (\ref{Eq:critDimitri}) here is sufficient for our general study.

Attaining the value of $\omega_{\rm crit}$ may be much easier for small bodies in white dwarf systems than in the solar system because the effective initial value of $\omega$ could be much higher around evolved stars. The reason is that during the highly luminous giant branch phases of evolution, YORP spin-up is ``supercharged" \citep{verjacgan2014,versch2020}. Hence the resulting spin rates of surviving asteroids around white dwarfs are likely to be, as a population, much more spun up toward the spin barrier than the asteroids of the current solar system population \citep{prahar2000}. Comets much further away than $\approx 100$~au throughout the main-sequence and giant branch phases are not likely to be as affected by strong radiative forces \citep{vereggganGB2015,verhigida2019}, and hence may not show enhanced initial spin rates around white dwarfs.

\subsection{YORP spin-up}\label{sec:YORP}

When comparing the effect of sublimative torques to the YORP effect, one potentially usable formulation which is general enough for exoplanetary systems is from \cite{verbirzam2022}, adapted from \cite{scheeres2007,scheeres2018}:

\[
\left(\frac{d\omega}{dt}\right)_{\rm YORP}
=
\frac{\chi \Phi R L_{\rm WD}}{m a^2 \sqrt{1-e^2} L_{\odot}}
\]
\begin{equation}
\ \ \ \ \ \ \ \ \ \ \ \ \ \ \ \ \ \ \ \,  \approx
\frac{3\chi \Phi L_{\rm WD}}{4 \pi \rho R^2 a^2 \sqrt{1-e^2} L_{\odot}},
\label{Eq:yorpeq}
\end{equation}

\noindent{}where $\Phi=1 \times 10^{17}$kg\,m\,s$^{-2}$ is the Solar radiation constant, $\chi$ is {\rev the aforementioned} dimensionless constant which incorporates the obliquity and asymmetry of the small object, and $a$ and $e$ are the orbital semi-major axis and eccentricity of the small body. Although many variables in equation (\ref{Eq:yorpeq}) are technically dependent on time, several may be considered fixed depending on the particular setup and what allows for analytical tractability. In this study, we set $e=0$.

\subsection{Relation with white dwarf properties}\label{sec:WDAge}

To self-consistently treat the value of $T$ from equation (\ref{Eq:SubSpin}) and the value of $L_{\rm WD}$ from equation (\ref{Eq:yorpeq}), as well as link these quantities with the age of the white dwarf, we need to identify relations between these quantities. Further, the MaMOS model outputs the intensity of starlight on the small body's surface, $I$, along with $T$, in a 1-to-1 mapping, requiring us to establish a link between $I$ and $L_{\rm WD}$.

First, we relate age and $L_{\rm WD}$. We use the term ``age" as shorthand for what is commonly known as the white dwarf's cooling age, which refers to the time elapsed since the star became a white dwarf. A compact relation between the white dwarf's age and effective temperature, $T_{\rm WD}$, can be roughly approximated by assuming a fiducial value of $M_{\rm WD}=0.60M_{\odot}$ with \citep{mcdver2021}
\[
\log_{10}\left(\frac{\rm Age}{\rm yr}\right) \approx 7.45 + 3.67 \log_{10}\left(\frac{T_{\rm WD}}{\rm K}\right) - 0.83 \left[\log_{10}\left(\frac{T_{\rm WD}}{\rm K}\right)\right]^2.
\]
\begin{equation}
\end{equation}
\noindent{}Also,
\begin{equation}
L_{\rm WD} = 4 \pi R_{\rm WD}^2 \varsigma_{\rm SB} T_{\rm WD}^4, 
\label{Eq:L4T}
\end{equation}
\noindent{}where $R_{\rm WD}$ is the radius of the white dwarf, and we assume that the emissivity of the small body surface equals unity. We compute $R_{\rm WD}$ by using our fiducial value of $M_{\rm WD}=0.60M_{\odot}$ and applying it to \citep{nauenberg1972}
\begin{equation}
\frac{R_{\rm WD}}{R_{\odot}}
\approx
0.0127 \left( \frac{M_{\rm WD}}{M_{\odot}} \right)^{-1/3}
\sqrt{1-0.607\left( \frac{M_{\rm WD}}{M_{\odot}} \right)^{4/3}},
\label{Eq:MR}
\end{equation}
\noindent{}yielding $R_{\rm WD} =$ 8,750~km. Finally,
\begin{equation}
I = \frac{L_{\rm WD}}{4\pi r^2}
\end{equation}
\noindent{}where
\begin{equation}
r = \frac{a\left(1-e^2\right)}{1+e \cos{f}}
\end{equation}
\noindent{}such that $f$ is the true anomaly of the small body orbit. For circular orbits, $r$ becomes a constant, independent of $f$.

\subsection{Application of MaMOS, SYORP, and YORP models}\label{sec:apply}

\begin{figure*}
\includegraphics[width=17.5cm]{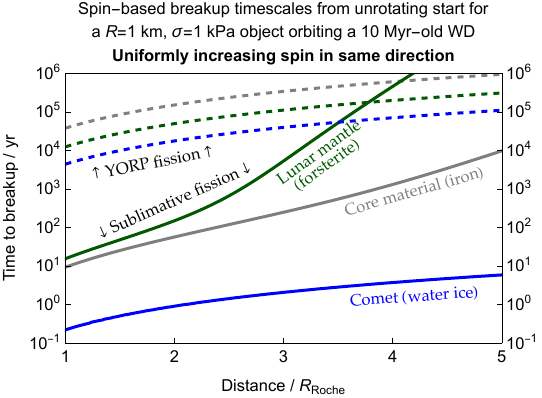}
\caption{
The time taken ($y$-axes) for sublimative outgassing (solid lines) and radiative torques (dashed lines) of the materials indicated to spin up an object with $R=1$~km from $\omega(t=0)=0$ to breakup speed. {\revv The spin-up is uniform and in the same direction}. The object is assumed to be on a circular or near-circular orbit located at a distance given on the $x$-axis, assuming that $R_{\rm Roche} = R_{\odot}$. The object has an internal tensile strength of $\sigma = 1$~kPa, and the cooling age of the white dwarf is 10~Myr. The figure illustrates that in this regime of parameter space, water ice outgassing is more important than YORP torques, and can break apart the object in under 10~yr; for monotonic spin-up, the breakup could occur even at a distance of $5R_{\rm Roche}$. A completely ``dry" object with this specific radius and internal strength would be subject to sublimative fission by iron or forsterite outgassing \rev only after 10~yr. 
}
\label{Fig:fiducial}
\end{figure*}

\begin{figure*}
\includegraphics[width=17.5cm]{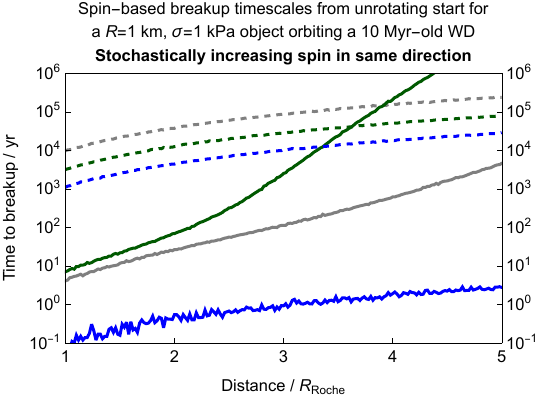}
\caption{
{\revv Same as Fig. \ref{Fig:fiducial}, except that the magnitude of the spin kicks is stochastic. For each curve, 200 different distances are sampled.}
}
\label{Fig:rand}
\end{figure*}

\begin{figure*}
\includegraphics[width=17.5cm]{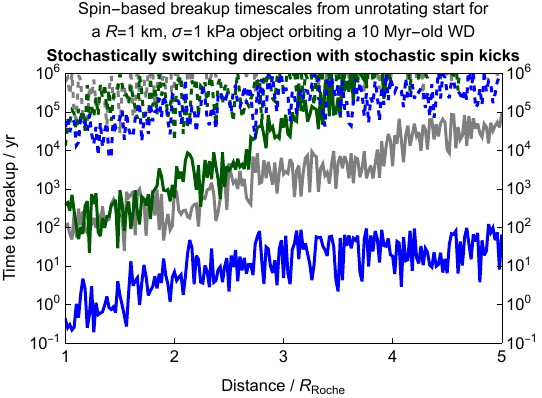}
\caption{
{\revv Same as Fig. \ref{Fig:rand}, except that the direction of the spin kick is now also stochastic.}
}
\label{Fig:signrand}
\end{figure*}

\begin{figure*}
\includegraphics[width=18.0cm]{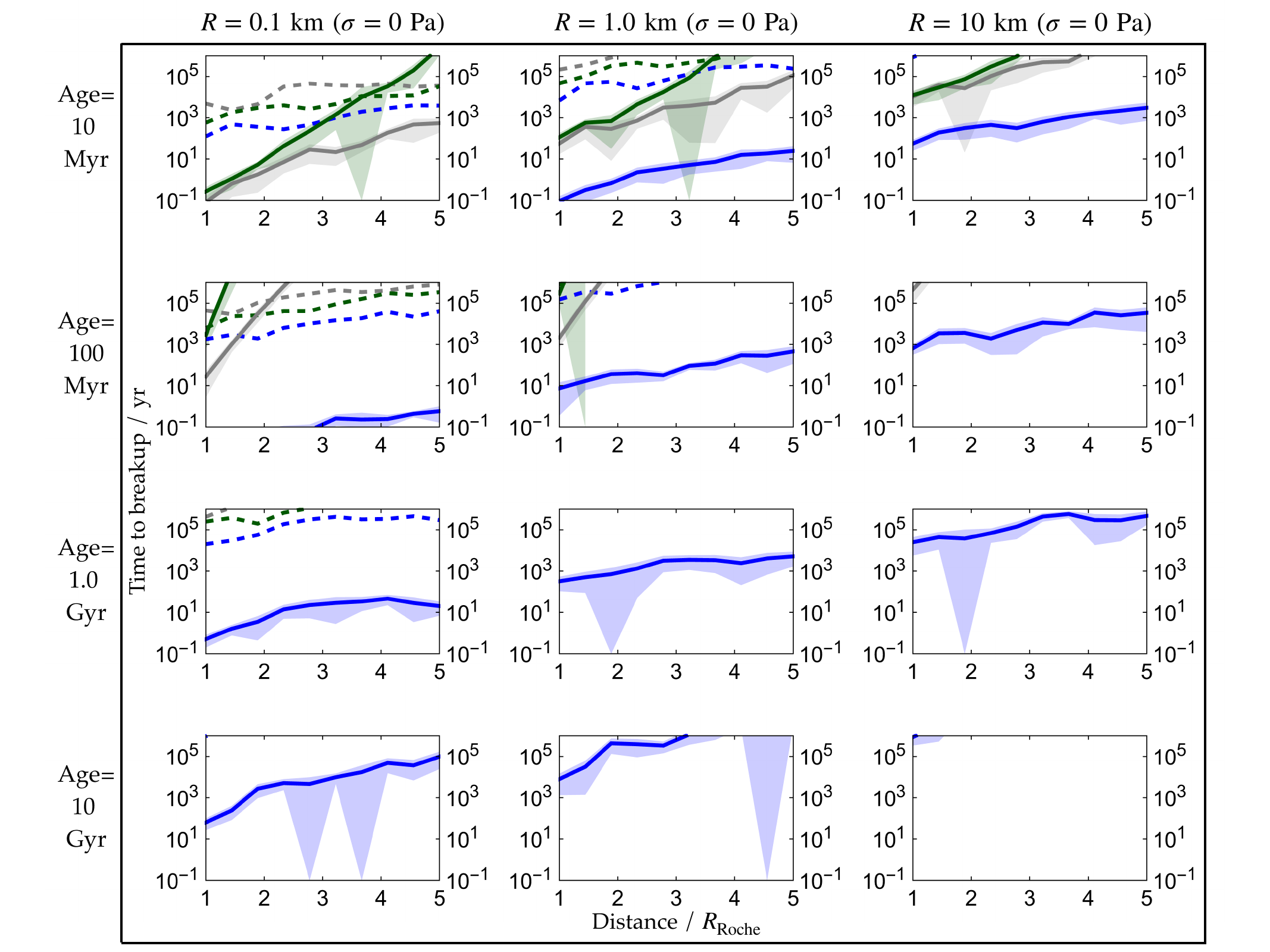}
\caption{
Like {\revv Fig. \ref{Fig:signrand}}, with exactly the same individual plot axes, but for different white dwarf cooling ages and object radii, and for strength-less objects. {\revv Here mean values of breakup timescales are shown as thick lines. For the SYORP curves only, the shaded regions correspond to one standard deviation; in cases where the standard deviation exceeds the mean, the region is drawn down to the bottom axis of the plot.} On all of the plots, the blue (water ice) solid curve resides below the other two solid curves, even when off of the plots. Further, the blue (water ice) solid curve always lies below the blue dashed (YORP) curve. This figure demonstrates that sublimative fission is primary relevant for objects with $R < 10$~km, and that for the oldest white dwarfs (10~Gyr), only water ice sublimative fission is plausible. For the youngest white dwarfs (10~Myr) and smallest objects, sublimative outgassing via iron or forsterite is likely for dry (volatile-poor) compositions.
}
\label{Fig:multi}
\end{figure*}

Having now established the necessary relations between the variables, we apply the MaMOS, SYORP, and YORP models and generate outputs. Figs. \ref{Fig:fiducial}-\ref{Fig:multi} illustrate our results, which focus on destruction timescales. {\revv Figs. \ref{Fig:fiducial}-\ref{Fig:signrand} represent our fiducial illustrations, whereas Fig. \ref{Fig:multi} displays a parameter sweep where each plot contains the same values on the $x$-axis and $y$-{\rev axes} as Figs. \ref{Fig:fiducial}-\ref{Fig:signrand}. In all of these plots, the curves represent the time required to spin up the small body from $\omega = 0$ to $\omega = \omega_{\rm crit}$}. 

{\revv
Figs. \ref{Fig:fiducial}-\ref{Fig:multi} display  different methods by which SYORP and YORP spin changes occur. In Fig. \ref{Fig:fiducial}, the magnitude of the spin kicks is uniform and in the same direction. In Fig. \ref{Fig:rand}, the magnitude of the spin kicks is stochastic, but still in the same direction. In Figs. \ref{Fig:signrand}-\ref{Fig:multi}, the magnitude of the spin kicks is stochastic, and the direction of the kick is selected stochastically. 
}

%%%%%%%%%%%%%%%% Figure
%%%%%%%%%%%%%%%% Figure

%%%%%%%%%%%%%%%% Figure
%%%%%%%%%%%%%%%% Figure

%%%%%%%%%%%%%%%% Figure
%%%%%%%%%%%%%%%% Figure

%%%%%%%%%%%%%%%% Figure
%%%%%%%%%%%%%%%% Figure

{\revv
We model spin changes through the treatment of the parameters $C_{\rm S}$ and $\chi$ at each timestep. When values for these parameters are selected stochastically, they are chosen from the physically motivated ranges $C_{\rm S} = 10^{-5}-10^{-3}$ (Section \ref{CSScale}) and $\chi=10^{-3}-10^{-2}$ \citep{scheeres2018}; in the uniform case, we adopted $C_{\rm S}=10^{-4}$ and $\chi=10^{-3}$. Because of the many orders-of-magnitude difference in timescales on these plots, we needed to stagger our choice of timesteps $\Delta t$: We first sampled $10^4$ uniform steps with $\Delta t = 10^{-2}$~yr, and then $10^4$ uniform steps with $\Delta t = 10^{-1}$~yr, and so forth, until we detected break-up for each curve.
}

{\revv
Unlike in Figs. \ref{Fig:fiducial}-\ref{Fig:signrand}, in Fig. \ref{Fig:multi} we plot only the mean values of the spin-up timescales, along with their one standard deviation variations in shaded regions. To obtain these statistical parameters, we ran 10 simulations at each distance from the white dwarf, for 10 different distances (in Figs. \ref{Fig:fiducial}-\ref{Fig:signrand} we ran 1 simulation at 200 different distances). In Fig. \ref{Fig:multi}, the shaded regions are shown only for the SYORP curves, for clarity; the downward spikes in the shaded regions indicate where the standard deviation is greater than the mean.
}

%Note that the plots are not duplicates, as in Fig. \ref{Fig:fiducial}, $\sigma = 1$~kPa, whereas in Fig. \ref{Fig:multi}, the small body is strength-less ($\sigma = 0$). 

%sf {\rev for constant values of $C_{\rm S}$ and $\chi$ (represented on the left $y$-axis) and for the population-level factor of 10 increase \citep{beretal2026} in case these values are stochastic (represented on the right $y$-axis)}. In this respect, we are illustrating conservatively long destruction timescales; the small bodies might arrive in the orbits already spinning, and perhaps close to $\omega_{\rm crit}$. {\rev Further, the upper bound for $C_{\rm S}$ is not reflected on this plot; adopting that would have decreased the values on the left $y$-axes by a factor of 10.} Hence, sublimative breakup could occur around even dimmer and colder white dwarfs. 

The $x$-axes on all of the plots display Roche radius, $R_{\rm Roche}$, assuming that $R_{\rm Roche} = R_{\odot}$ (representing a typical estimate; see \citealt{vercarleietal2017} for variations based on physical properties of the small object). We chose the $x$-axis range to roughly correspond to orbital periods of 5-25 hours, depending on the white dwarf mass (see Fig. 1 of \citealt*{vercuk2025}). Further, as previously mentioned, in all plots we assume that $e=0$, partly motivated by the {\revv orbit circularisation and shrinkage predictions in} \cite{lietal2025a}, and partly motivated by considering orbits whose pericentres exceed the tidal disruption limit. Furthermore, exploring significantly eccentric orbits would lead to an overly unconstrained parameter space for the purposes of this initial study, and are therefore not considered.

On all of the plots, the solid lines refer to sublimative fission timescales, and the dashed lines refer to the YORP fission timescales. The colour-coding corresponds to the materials specified with annotations in Fig. \ref{Fig:fiducial}. 
%Throughout the plot domains, the vertical ordering of the curves for sublimative fission is always maintained, including some curves that are outside of the plotted area. The same is true of the YORP curves.

Now focussing in {\revv on Fig. \ref{Fig:fiducial}-\ref{Fig:signrand}}, we consider the destruction timescales for a $R=1$~km small body that has some internal tensile strength ($\sigma = 1$~kPa) orbiting a star that has been a white dwarf for 10 Myr. The {\revv plots illustrate that regardless of the time dependence on spin}, just at the Roche radius, any of water ice, iron or forsterite could generate sublimative fission within, at most, 10~yr, depending on which materials are prevalent. In all cases, at $r=1R_{\rm Roche}$, the sublimative destruction timescale is shorter than the YORP timescale. 
%If water ice is subliming, then the object would disrupt in about 0.2~yr.
The YORP and sublimation curves begin to overlap only {\revv at a distance of $r\approx3-4 R_{\rm Roche}$}, and only for forsterite. In contrast, sublimative fission due to water ice outgassing occurs more quickly than YORP fission throughout the $r = 1-5 R_{\rm Roche}$ range.

{\revv
The differences amongst Figs. \ref{Fig:fiducial}-\ref{Fig:signrand} are starkest when the direction of the spin changes stochastically. The spin-up timescales in Fig. \ref{Fig:signrand} are roughly one order of magnitude higher than those in Figs. \ref{Fig:fiducial} and \ref{Fig:rand}. This difference is equal to that which was empirically deduced from solar system YORP observations in \cite{beretal2026} when they compared the observations to a theoretical uniform increase in spin in the same direction.
}

Because white dwarfs with an age of $\sim 10$~Myr (as in Fig. \ref{Fig:fiducial}) are the youngest that have been observed to host planetary systems \citep[][]{ganetal2019}, we are taking a snapshot of the extreme end of the observable population. {\rev Nevertheless,} we also need to consider older white dwarfs. In fact, on the opposite extreme, planetary metals have been observed in stars which have been white dwarfs for $\sim 10$~Gyr \citep{elmetal2022}; at this age, the residence timescale of the metals in the white dwarf photosphere is effectively negligible, being 4-13 orders of magnitude shorter than 10~Gyr \citep{koester2009}.

Hence, in Fig. \ref{Fig:multi}, we explore sublimative fission across all relevant white dwarf ages, and for three different small body radii (spin rate evolution has a strong functional dependence on radius, as seen in equations \ref{Eq:SubSpin} and \ref{Eq:yorpeq}).
%(which has a strong functional dependence on spin rate, as seen in equations \ref{Eq:SubSpin} and \ref{Eq:yorpeq}). 
{\rev These radii were chosen to encompass the parameter space that is relevant to SYORP, and because debris orbiting white dwarfs is thought to range from micron-sized dust to objects which are nearly as large as Ceres \citep{rapetal2016,guretal2017}.}

All of the plots in this figure model strength-less small bodies, an assumption often adopted for white dwarf planetary systems given the prevalence of rubble piles in the solar system. Comparison between Figs. \ref{Fig:signrand} and the top middle plot of \ref{Fig:multi} reveals {\revv a slight difference in} the destruction timescales when reducing the tensile strength from $\sigma=1$~kPa to zero; the rubble piles break up more quickly. 

Fig. \ref{Fig:multi} also indicates that white dwarfs with ages of $\sim 100$~Myr can easily effectuate water ice-driven sublimative fission for $\lesssim 1$~km objects, as can $1$~Gyr-old white dwarfs for $\lesssim 0.1$~km objects. The oldest known metal-enriched white dwarfs ($10$~Gyr) are most likely too dim to generate sublimative fission for objects larger than boulders unless they are already spinning close to $\omega_{\rm crit}$.

The figure also suggests that sublimation should not exclusively be confined to traditional volatiles, such as water and CO$_2$; completely dehydrated, iron-rich planetary fragments around young white dwarfs can also be spun up to fission through sublimation. In fact, the upper left plot in Fig. \ref{Fig:multi} displays multiple plausible pathways to rotational fission, perhaps suggesting that debris discs might form more quickly in this area of parameter space.

Fig. \ref{Fig:multi} additionally hints at the maximum value of $R$ for which sublimative fission could occur. Since the discovery of the first transiting debris \citep{vanetal2015}, about one decade has now passed. Throughout this time, Fig. \ref{Fig:multi} suggests that we have not seen sublimative fission (from a standing start) of an object greater than about 10~km in size.

{\rev Finally, we can contexutalise these results with respect to these systems' snowlines, $r_{\rm sub}$ (often referred to as ``sublimation radii” in the white dwarf planetary system literature). These snowlines are dependent on the white dwarf cooling age, $t_{\rm cool}$, as well as the sublimation temperature, $T_{\rm sub}$. One formulation that can estimate $r_{\rm sub}$ is equation 67 of \cite{veridagrietal2023}:

\begin{equation}
r_{\rm sub} = \frac{1}{4}
              \sqrt{\frac{3.53 L_{\odot}}{\pi \sigma T_{\rm sub}^4}
                    \left( \frac{M_{\star}}{0.65 M_{\odot}} \right)  
                    \left(0.1 + \frac{t_{\rm cool}}{\rm Myr} \right)^{-1.18} }.
\label{rsub}
\end{equation}

\noindent{}This equation combines relations from \cite{mestel1952} and \cite{rafikov2011a}, and is less accurate for white dwarfs which have already experienced crystallization (likely at $t_{\rm cool} \approx 10$~Gyr). Given $T_{\rm sub} = \left\lbrace 2160~{\rm K}, 2560~{\rm K}, 273~{\rm K} \right\rbrace$ for iron, forsterite, and water ice, respectively \citep{steckloffetal2021b}, we obtain, for $t_{\rm cool} = 10~{\rm Myr}$, $ r_{\rm sub} \approx \left\lbrace 1.6, 1.2, 100 \right\rbrace R_{\odot}$; 
for $t_{\rm cool} = 100~{\rm Myr}$, $r_{\rm sub} \approx \left\lbrace 0.4, 0.3, 27 \right\rbrace R_{\odot}$;
for $t_{\rm cool} = 1~{\rm Gyr}$, $r_{\rm sub} \approx \left\lbrace 0.1, 0.08, 7 \right\rbrace R_{\odot}$; 
for $t_{\rm cool} = 10~{\rm Gyr}$, $ r_{\rm sub} \approx \left\lbrace 0.03, 0.02, 2  \right\rbrace R_{\odot}$. 
}

\section{Discussion}\label{sec:discussion}

\subsection{Robustness of our results}

Because Figs. \ref{Fig:fiducial}-\ref{Fig:multi} assume fixed values for several variables, one may wonder how the qualitative results of this study hold up when different parameters are adopted for these variables. We now consider these in turn:

\begin{enumerate}

\item $M_{\rm WD}, R_{\rm WD}$: The assumption $M_{\rm WD}=0.6M_{\odot}$ has a negligible effect on our results. The vast majority of observed single white dwarfs have masses in the range $0.4M_{\odot}-0.8M_{\odot}$ \citep[e.g.][]{treetal2016,cunetal2024,obretal2024}, which lead, through equation (\ref{Eq:MR}), to a radius range of $R_{\rm WD}\approx 7,000-11,000$~km, representing up to a 25 per\,cent difference from the nominal $R_{\rm WD}=$~8,750~km value. Then, equation (\ref{Eq:L4T}) shows that the corresponding temperature ($T_{\rm WD}$) difference would be just 10-15 per\,cent, altering our timescale estimates by the same fraction.

\item $\alpha$: The value of the sublimative spin rate is a linear function of $\alpha$ (equation \ref{Eq:SubSpin}), and representative ranges of $\alpha$ are displayed in Table \ref{TableInParam_models} (applicable to water ice and forsterite only, as for iron, $\alpha = 1$). Hence, we could have reasonably doubled or reduced by two-thirds our adopted values of $\alpha_{\rm water \ ice}=0.50$ or $\alpha_{\rm forsterite}=0.10$. Doing so would have changed the destruction timescales by an equal measure.

\item $C_{\rm S}$: Similar to $\alpha$, {\rev as already discussed, }the value of the sublimative spin rate is a linear function of $C_{\rm S}$ (equation \ref{Eq:SubSpin}), and the most representative ranges of $C_{\rm S}$ are displayed in Table \ref{TableInParam_models}. 
%The value of $C_{\rm S}$ could have been raised or lowered by one order of magnitude, with a corresponding change in destruction timescale.

\item $\chi$: The variable $\chi$ affects only the YORP spin-up timescale, and linearly so (equation \ref{Eq:spinsub}), and physically conveys that the small object needs to have some asymmetry for the YORP effect to initiate. Hence, a perfectly spherical object will have $\chi=0$, and in this case, the YORP effect would be a nonfactor, eliminating the dashed curves on all of our plots. In the other extreme, \cite{scheeres2018} adopted an upper bound of $\chi=10^{-2}$, which is one order of magnitude higher than our adopted value. This higher value would lower all of the YORP curves on our plots by one order of magnitude, primarily affecting the results for white dwarfs younger than about 100~Myr.
    
\end{enumerate}

\subsection{Retention of subliming material}

Before reaching the close vicinity of the white dwarf, the small object may lose its sublimation-susceptible material during the precursor giant branch phases of stellar evolution (where max($L$) = $10^3-10^4L_{\odot}$). A related focus of previous studies has been the retention of specifically {\it volatiles} in the interiors of these objects during these stellar phases \citep{jurxu2010,juretal2012,malper2016,malper2017a,malper2017b}, which illustrate that as long as a small object is sufficiently far from the giant branch star, then the object can retain its subliming material. Subsequently, after the star has become a white dwarf, the small objects can be gravitationally scattered inwards towards the white dwarf through a variety of means (see \citealt*{veras2024} for a recent review of mechanisms and references).

A simple order-of-magnitude calculation reveals how far into the surfaces of distant objects the thermal pulse can penetrate over the duration of the (Asymptotic Giant Branch) AGB phase, which typically lasts $\sim$0.1-1 Myr. Let the thermal skin depth $d_{\rm therm}$ represent the depth over which the amplitude of the thermal wave drops to 1/2.72 of its value at the surface \cite[e.g.,][]{steckloffetal2021a}. The thermal pulse rapidly damps out over a depth of several thermal skin depths, which is related to both the thermal diffusivity of the material ($\mathcal{H}$) and the duration of the thermal pulse, $\tau_{\rm pulse}$, as
\begin{equation}
    d_{\rm therm} = \sqrt{\mathcal{H}\tau_{\rm pulse}}.
\end{equation}

For typical cometary materials, the thermal diffusivity is on the order of $\mathcal{H}_{\rm comet}\sim10^{-8}-10^{-7}$ m$^2$/s \citep{steckloffetal2021a}, although for consolidated rock $\mathcal{H}_{\rm rock}\sim10^{-6}$ m$^2$/s, and for iron  $\mathcal{H}_{\rm iron}\sim10^{-5}$ m$^2$/s.
By applying these values, we find that even highly volatile-rich icy cometary bodies would only lose volatiles from the upper few km of their surfaces. More consolidated rock, such as mantle material, could be thermally processed in the upper $\sim$1-10 km of their surfaces. However, the high latent heat of sublimation of rock would limit how much of this material would ultimately be vaporized \citep{yuqilietal2024}. 

Iron bodies represent a unique middle case, where they have high thermal diffusivity and intermediate volatility (i.e., latent heat of sublimation). Iron objects could be processed throughout ($\sim$10-100 km deep), and are typically located within the water ice snowline. Thus, during the AGB phase, bare iron objects may be preferentially cleared from the system.  Nevertheless, were the AGB evolution to dynamically stir the small body populations of the host star, then the resulting collisional evolution could break up dwarf-planet-sized objects (e.g,, Vesta and Ceres), leading to the potential replenishment of this population. Alternatively, iron-rich bodies may be sufficiently impure as to build up lag deposits of more refractory materials on their surfaces, leading to a projective mantle forming.

A related question is: does sublimative fission primarily occur on objects scattered into the close vicinity of white dwarfs, or primarily on their fragments after the progenitor body has already broken up through some other process? The answer is probably both, given the applicable size regime (up to $R \approx 10$~km) demonstrated here. Small objects survive the post-scattered journey to the close vicinity of the white dwarf unless they are in the cm-to-m size regime \citep{broetal2017,mcdver2021}, when complete sublimation can occur.

\subsection{Repeated sublimative fissions}

An as-yet-unexplained feature of two of the white dwarf planetary systems which feature transiting debris is why the {\revv photometric} activity levels in those systems have decreased or disappeared entirely. The initially high level of activity in WD~1145+017 vanished in just under 10 years \citep{aunetal2024}, whereas in ZTF~J1944+4557, the activity ceased in about just two years \citep{guietal2025}.

{\revv In both cases, the cessation of activity is indicated by the measured periodicity disappearing from the transit curve, and the transit curve becoming flat. This disappearance of the periodicity is inferred to represent the destruction of a minor planet into dust. Why the periodicities disappear on these yearly to decadal timescales remains unanswered.}

Hence, understanding the fate of rotationally fissioned fragments is important. In this subsection, we construct a basic upper bound for the time required for successive fissions to grind down the progenitor object into dust. This estimate represents an upper bound because small fragments of the parent body may be sublimated, obviating successive grind-downs.

{\revv 
The size threshold below which direct sublimation outpaces further fission may be approximated by combining equations 17 and 18 of \citep{mcdver2021} and can be expressed as $R_{\rm WD}^{5/2} T_{\rm WD}^4 \varsigma_{SB}/(2\sqrt{2GM_{\rm WD}} \lambda \rho r)$. Hence, for a fixed white dwarf, this threshold is dependent on the cooling age of the white dwarf, the separation from the white dwarf, and the density and latent heat of sublimation of the material. The size threshold for \{ iron, forsterite, water ice\} equals, for cooling ages of 10~Myr, $\left\lbrace 0.08{\rm m}-0.4{\rm m}, 0.1{\rm m}-0.6{\rm m}, 4{\rm m}-20{\rm m} \right\rbrace$ and for cooling ages of 100 Myr, $ \left\lbrace 0.006{\rm m}-0.03{\rm m}, 0.009{\rm m}-0.05{\rm m}, 0.3{\rm m}-1.6{\rm m} \right\rbrace$, where the ranges correspond to distances of $1-5 R_{\rm Roche}$.

}

We denote the sublimative {\revv spin-up} destruction timescale for a single fission as $\tau_{\rm SYORP}$. Fissions likely create a small number of fragments (nominally two), each spinning at the same rotation speed as before fragmentation; the difference in total rotational kinetic energy accounts for the translational kinetic energy of the fragments relative to one another. Further, for a strength-dominated object made of real geologic material, its strength actually increases as it gets smaller, due to the reduction in the number of flaws in the remaining object \citep{brace1961,stejac2016}. Nevertheless, the reduction in the destruction timescale is dominated by the reduction in $R$ rather than the increase in $\sigma$.

First, let us assume that, upon disruption, the object of radius $R_0$ fails along a plane of weakness to produce two equal-mass halves each of radius $R_1$. Hence, in order to conserve mass/volume, 
\begin{equation}
R_1 = \left(\frac{1}{2}\right)^{1/3}R_0.
\end{equation}
\noindent{}More generically, the size of the $R_{n+1}$th fragment is always going to be related to the size of the $R_n$th fragment via
\begin{equation}
R_{n+1} = \left(\frac{1}{2}\right)^{1/3}R_n,
\end{equation} 
\noindent{}with each fragmentation requiring the object to spin up to breakup speed over this timescale. 

In order to obtain our upper bound, we can now further assume monotonic spin-up as opposed to stochastic spin-up. Hence, the total amount of time a progenitor small body will take to be ground down into dust through a succession of sublimative fissions is (assuming the fragments do not first sublimate into gas)
\begin{equation}
    \tau_{\rm tot} = \sum_{n=0}^{\infty} \left( \frac{1}{2} \right)^{-\frac{2n}{3}}\tau_{\rm SYORP} \approx 2.7 \tau_{\rm SYORP}.
\end{equation}
Given the order-of-magnitude nature of this investigation, the difference between $1.0\tau_{\rm SYORP}$ and $2.7\tau_{\rm SYORP}$ is effectively indistinguishable, and $2.7\tau_{\rm SYORP}$ represents just an upper bound anyway ({\revv incorporating the size threshold where fragments are completely sublimated would reduce this coefficient to a number between 1.0 and 2.7}). Hence, to a good approximation, we may assume that the timescale for sublimative fission is the same timescale to reduce the fragments into dust or gas.

%A caveat is that this model tacitly assumes that sublimative torques only act in one direction (i.e., that torques only spin the object up).  In reality, sublimative torques are likely to be analogous to the YORP effect, with the magnitudes and directions of these torques being so sensitive to shape that the new shapes of the fragments have an equal chance of ending up with a configuration that causes further spin up or spin down to a more stable spin period \citep{schetal2008,statler2009}. 

%Even without disrupting, comet nuclei can change their activity patterns via mass wasting \citep{steetal2016}. Additionally, the shapes of real comet nuclei can experience sublimative torques whose magnitudes and directions are quite sensitive to the subsolar latitude \citep{hiretal2016}. Due to this stochasticity, a more realistic timescale for fragmentation to dust is likely to be much longer than this simplified model predicts \citep{cotetal2015,hiretal2016}. Nevertheless, our model provides a general timescale and minimum constraint on this process, with the actual timescale of this process being perhaps an order of magnitude longer than $2.7\tau_{\rm SYORP}$.

\section{Summary}\label{sec:summary}

Observations demonstrate that planetary debris orbit white dwarfs at distances of $r\approx1-4R_{\rm Roche}$, indicating that tidal disruption of minor planets within the Roche sphere may not be as prevalent as previously assumed. Here, we explore the alternate disruption mechanism of rotational fission due to sublimitive outgassing, and in particular for water ice, iron and forsterite. We show that this process easily breaks apart small asteroids, comets and fragments exterior to the Roche sphere of white dwarfs with cooling ages of up to 100~Myr, and older in specific cases. We also demonstrate that sublimative fission is usually more efficient than breakup due to the YORP effect, especially for water ice outgassing, and hence could play a significant role in explaining the evolution of transiting debris around young white dwarfs.

\section*{Acknowledgements}

{\rev We thank the expert reviewer for constructive comments that have improved the manuscript}. J.K.S. and K.V. were supported by NASA grant 80NSSC20K0267.

\section*{Data Availability}

All data presented in this paper is available upon reasonable request to the authors.

%%%%%%%%%%%%%%%%%%%% REFERENCES %%%%%%%%%%%%%%%%%%

\bibliographystyle{mnras}
\bibliography{DVbibfile13}

@ARTICLE{aunetal2024,
       author = {{Aungwerojwit}, Amornrat and {G{\"a}nsicke}, Boris T. and {Dhillon}, Vikram S. and {Drake}, Andrew and {Inight}, Keith and {Kaye}, Thomas G. and {Marsh}, T.~R. and {Mullen}, Ed and {Pelisoli}, Ingrid and {Swan}, Andrew},
        title = "{Long-term variability in debris transiting white dwarfs}",
      journal = {\mnras},
         year = 2024,
        month = may,
       volume = {530},
       number = {1},
        pages = {117-128},
          doi = {10.1093/mnras/stae750},
archivePrefix = {arXiv},
       eprint = {2404.04422},
 primaryClass = {astro-ph.EP},
       adsurl = {https://ui.adsabs.harvard.edu/abs/2024MNRAS.530..117A}
}

@ARTICLE{beretal2026,
       author = {{Bertinelli}, G. and {Zhou}, W.-H. and {Tanga}, P.},
        title = "{Exploring rotational properties and the YORP effect in asteroid families}",
      journal = {\aap},
         year = 2026,
        month = mar,
       volume = {707},
          eid = {A135},
        pages = {A135},
          doi = {10.1051/0004-6361/202558167},
archivePrefix = {arXiv},
       eprint = {2601.12972},
 primaryClass = {astro-ph.EP},
       adsurl = {https://ui.adsabs.harvard.edu/abs/2026A&A...707A.135B}
}

@ARTICLE{bhaetal2025,
       author = {{Bhattacharjee}, Soumyadeep and {Vanderbosch}, Zachary P. and {Hollands}, Mark A. and {Tremblay}, Pier-Emmanuel and {Xu}, Siyi and {Guidry}, Joseph A. and {Hermes}, J.~J. and {Caiazzo}, Ilaria and {Rodriguez}, Antonio C. and {van Roestel}, Jan and {Roulston}, Benjamin R. and {Riddle}, Reed and {Rusholme}, Ben and {Groom}, Steven L. and {Smith}, Roger and {Toloza}, Odette},
        title = "{A ZTF Search for Circumstellar Debris Transits in White Dwarfs: Six New Candidates, one with Gas Disk Emission, identified in a Novel Metric Space}",
      journal = {arXiv e-prints},
         year = 2025,
        month = feb,
          eid = {arXiv:2502.05502},
        pages = {arXiv:2502.05502},
          doi = {10.48550/arXiv.2502.05502},
archivePrefix = {arXiv},
       eprint = {2502.05502},
 primaryClass = {astro-ph.SR},
       adsurl = {https://ui.adsabs.harvard.edu/abs/2025arXiv250205502B}
}

@ARTICLE{blaetal2021,
       author = {{Blackman}, J.~W. and {Beaulieu}, J.~P. and {Bennett}, D.~P. and {Danielski}, C. and {Alard}, C. and {Cole}, A.~A. and {Vandorou}, A. and {Ranc}, C. and {Terry}, S.~K. and {Bhattacharya}, A. and {Bond}, I. and {Bachelet}, E. and {Veras}, D. and {Koshimoto}, N. and {Batista}, V. and {Marquette}, J.~B.},
        title = "{A Jovian analogue orbiting a white dwarf star}",
      journal = {\nat},
         year = 2021,
        month = oct,
       volume = {598},
       number = {7880},
        pages = {272-275},
          doi = {10.1038/s41586-021-03869-6},
archivePrefix = {arXiv},
       eprint = {2110.07934},
 primaryClass = {astro-ph.EP},
       adsurl = {https://ui.adsabs.harvard.edu/abs/2021Natur.598..272B}
}

@INPROCEEDINGS{bonsor2026,
       author = {{Bonsor}, Amy},
        title = "{White dwarf systems: The composition of exoplanets}",
    booktitle = {Encyclopedia of Astrophysics},
         year = 2026,
       volume = {1},
        month = jan,
        pages = {564-572},
          doi = {10.1016/B978-0-443-21439-4.00020-1},
       adsurl = {https://ui.adsabs.harvard.edu/abs/2026enap....1..564B}
}

@ARTICLE{botetal2006,
       author = {{Bottke}, Jr., William F. and {Vokrouhlick{\'y}}, David and {Rubincam}, David P. and {Nesvorn{\'y}}, David},
        title = "{The Yarkovsky and Yorp Effects: Implications for Asteroid Dynamics}",
      journal = {Annual Review of Earth and Planetary Sciences},
         year = 2006,
        month = may,
       volume = {34},
        pages = {157-191},
          doi = {10.1146/annurev.earth.34.031405.125154},
       adsurl = {https://ui.adsabs.harvard.edu/abs/2006AREPS..34..157B}
}

@ARTICLE{brace1961,
       author = {{Brace}, W.~F.},
        title = "{Mohr Construction in the Analysis of Large Geologic Strain}",
      journal = {Geological Society of America Bulletin},
         year = 1961,
        month = jan,
       volume = {72},
       number = {7},
        pages = {1059},
          doi = {10.1130/0016-7606(1961)72[1059:MCITAO]2.0.CO;2},
       adsurl = {https://ui.adsabs.harvard.edu/abs/1961GSAB...72.1059B}
}

@ARTICLE{broetal2017,
       author = {{Brown}, John C. and {Veras}, Dimitri and {G{\"a}nsicke}, Boris T.},
        title = "{Deposition of steeply infalling debris around white dwarf stars}",
      journal = {\mnras},
         year = 2017,
        month = jun,
       volume = {468},
       number = {2},
        pages = {1575-1593},
          doi = {10.1093/mnras/stx428},
archivePrefix = {arXiv},
       eprint = {1702.05109},
 primaryClass = {astro-ph.EP},
       adsurl = {https://ui.adsabs.harvard.edu/abs/2017MNRAS.468.1575B}
}

@ARTICLE{cunetal2024,
       author = {{Cunningham}, Tim and {Tremblay}, Pier-Emmanuel and {W. O'Brien}, Mairi},
        title = "{Initial-final mass relation from white dwarfs within 40 pc}",
      journal = {\mnras},
         year = 2024,
        month = jan,
       volume = {527},
       number = {2},
        pages = {3602-3611},
          doi = {10.1093/mnras/stad3275},
archivePrefix = {arXiv},
       eprint = {2310.15410},
 primaryClass = {astro-ph.SR},
       adsurl = {https://ui.adsabs.harvard.edu/abs/2024MNRAS.527.3602C}
}

@ARTICLE{denetal2020,
       author = {{Dennihy}, Erik and {Xu}, Siyi and {Lai}, Samuel and {Bonsor}, Amy and {Clemens}, J.~C. and {Dufour}, Patrick and {G{\"a}nsicke}, Boris T. and {Gentile Fusillo}, Nicola Pietro and {Hardy}, Fran{\c{c}}ois and {Hegedus}, R.~J. and {Hermes}, J.~J. and {Kaiser}, B.~C. and {Kissler-Patig}, Markus and {Klein}, Beth and {Manser}, Christopher J. and {Reding}, Joshua S.},
        title = "{Five New Post-main-sequence Debris Disks with Gaseous Emission}",
      journal = {\apj},
         year = 2020,
        month = dec,
       volume = {905},
       number = {1},
          eid = {5},
        pages = {5},
          doi = {10.3847/1538-4357/abc339},
archivePrefix = {arXiv},
       eprint = {2010.03693},
 primaryClass = {astro-ph.EP},
       adsurl = {https://ui.adsabs.harvard.edu/abs/2020ApJ...905....5D}
}

@ARTICLE{duvetal2020,
       author = {{Duvvuri}, Girish M. and {Redfield}, Seth and {Veras}, Dimitri},
        title = "{Necroplanetology: Simulating the Tidal Disruption of Differentiated Planetary Material Orbiting WD 1145+017}",
      journal = {\apj},
         year = 2020,
        month = apr,
       volume = {893},
       number = {2},
          eid = {166},
        pages = {166},
          doi = {10.3847/1538-4357/ab7fa0},
archivePrefix = {arXiv},
       eprint = {2003.08410},
 primaryClass = {astro-ph.EP},
       adsurl = {https://ui.adsabs.harvard.edu/abs/2020ApJ...893..166D}
}

@ARTICLE{elmetal2022,
       author = {{Elms}, Abbigail K. and {Tremblay}, Pier-Emmanuel and {G{\"a}nsicke}, Boris T. and {Koester}, Detlev and {Hollands}, Mark A. and {Gentile Fusillo}, Nicola Pietro and {Cunningham}, Tim and {Apps}, Kevin},
        title = "{Spectral analysis of ultra-cool white dwarfs polluted by planetary debris}",
      journal = {\mnras},
         year = 2022,
        month = dec,
       volume = {517},
       number = {3},
        pages = {4557-4574},
          doi = {10.1093/mnras/stac2908},
archivePrefix = {arXiv},
       eprint = {2206.05258},
 primaryClass = {astro-ph.SR},
       adsurl = {https://ui.adsabs.harvard.edu/abs/2022MNRAS.517.4557E}
}

@ARTICLE{faretal2009,
       author = {{Farihi}, J. and {Jura}, M. and {Zuckerman}, B.},
        title = "{Infrared Signatures of Disrupted Minor Planets at White Dwarfs}",
      journal = {\apj},
         year = 2009,
        month = apr,
       volume = {694},
       number = {2},
        pages = {805-819},
          doi = {10.1088/0004-637X/694/2/805},
archivePrefix = {arXiv},
       eprint = {0901.0973},
 primaryClass = {astro-ph.EP},
       adsurl = {https://ui.adsabs.harvard.edu/abs/2009ApJ...694..805F}
}

@ARTICLE{faretal2010,
       author = {{Farihi}, J. and {Jura}, M. and {Lee}, J. -E. and {Zuckerman}, B.},
        title = "{Strengthening the Case for Asteroidal Accretion: Evidence for Subtle and Diverse Disks at White Dwarfs}",
      journal = {\apj},
         year = 2010,
        month = may,
       volume = {714},
       number = {2},
        pages = {1386-1397},
          doi = {10.1088/0004-637X/714/2/1386},
archivePrefix = {arXiv},
       eprint = {1003.2627},
 primaryClass = {astro-ph.EP},
       adsurl = {https://ui.adsabs.harvard.edu/abs/2010ApJ...714.1386F}
}

@ARTICLE{faretal2013,
       author = {{Farihi}, J. and {G{\"a}nsicke}, B.~T. and {Koester}, D.},
        title = "{Evidence for Water in the Rocky Debris of a Disrupted Extrasolar Minor Planet}",
      journal = {Science},
         year = 2013,
        month = oct,
       volume = {342},
       number = {6155},
        pages = {218-220},
          doi = {10.1126/science.1239447},
archivePrefix = {arXiv},
       eprint = {1310.3269},
 primaryClass = {astro-ph.EP},
       adsurl = {https://ui.adsabs.harvard.edu/abs/2013Sci...342..218F}
}

@ARTICLE{faretal2022,
       author = {{Farihi}, J. and {Hermes}, J.~J. and {Marsh}, T.~R. and {Mustill}, A.~J. and {Wyatt}, M.~C. and {Guidry}, J.~A. and {Wilson}, T.~G. and {Redfield}, S. and {Izquierdo}, P. and {Toloza}, O. and {G{\"a}nsicke}, B.~T. and {Aungwerojwit}, A. and {Kaewmanee}, C. and {Dhillon}, V.~S. and {Swan}, A.},
        title = "{Relentless and complex transits from a planetesimal debris disc}",
      journal = {\mnras},
         year = 2022,
        month = apr,
       volume = {511},
       number = {2},
        pages = {1647-1666},
          doi = {10.1093/mnras/stab3475},
archivePrefix = {arXiv},
       eprint = {2109.06183},
 primaryClass = {astro-ph.EP},
       adsurl = {https://ui.adsabs.harvard.edu/abs/2022MNRAS.511.1647F}
}

@ARTICLE{faretal2025,
       author = {{Farihi}, J. and {Su}, K.~Y.~L. and {Melis}, C. and {Kenyon}, S.~J. and {Swan}, A. and {Redfield}, S. and {Wyatt}, M.~C. and {Debes}, J.~H.},
        title = "{Subtle and Spectacular: Diverse White Dwarf Debris Disks Revealed by JWST}",
      journal = {\apjl},
         year = 2025,
        month = mar,
       volume = {981},
       number = {1},
          eid = {L5},
        pages = {L5},
          doi = {10.3847/2041-8213/adae88},
archivePrefix = {arXiv},
       eprint = {2501.18338},
 primaryClass = {astro-ph.EP},
       adsurl = {https://ui.adsabs.harvard.edu/abs/2025ApJ...981L...5F}
}

@ARTICLE{genetal2021,
       author = {{Gentile Fusillo}, N.~P. and {Manser}, C.~J. and {G{\"a}nsicke}, Boris T. and {Toloza}, O. and {Koester}, D. and {Dennihy}, E. and {Brown}, W.~R. and {Farihi}, J. and {Hollands}, M.~A. and {Hoskin}, M.~J. and {Izquierdo}, P. and {Kinnear}, T. and {Marsh}, T.~R. and {Santamar{\'\i}a-Miranda}, A. and {Pala}, A.~F. and {Redfield}, S. and {Rodr{\'\i}guez-Gil}, P. and {Schreiber}, M.~R. and {Veras}, Dimitri and {Wilson}, D.~J.},
        title = "{White dwarfs with planetary remnants in the era of Gaia - I. Six emission line systems}",
      journal = {\mnras},
         year = 2021,
        month = jun,
       volume = {504},
       number = {2},
        pages = {2707-2726},
          doi = {10.1093/mnras/stab992},
archivePrefix = {arXiv},
       eprint = {2010.13807},
 primaryClass = {astro-ph.SR},
       adsurl = {https://ui.adsabs.harvard.edu/abs/2021MNRAS.504.2707G}
}

@ARTICLE{graetal1990,
       author = {{Graham}, James R. and {Matthews}, K. and {Neugebauer}, G. and {Soifer}, B.~T.},
        title = "{The Infrared Excess of G29--38: A Brown Dwarf or Dust?}",
      journal = {\apj},
         year = 1990,
        month = jul,
       volume = {357},
        pages = {216},
          doi = {10.1086/168907},
       adsurl = {https://ui.adsabs.harvard.edu/abs/1990ApJ...357..216G}
}

@ARTICLE{guietal2021,
       author = {{Guidry}, Joseph A. and {Vanderbosch}, Zachary P. and {Hermes}, J.~J. and {Barlow}, Brad N. and {Lopez}, Isaac D. and {Boudreaux}, Thomas M. and {Corcoran}, Kyle A. and {Bell}, Keaton J. and {Montgomery}, M.~H. and {Heintz}, Tyler M. and {Castanheira}, Barbara G. and {Reding}, Joshua S. and {Dunlap}, Bart H. and {Winget}, D.~E. and {Winget}, Karen I. and {Kuehne}, J.~W.},
        title = "{I Spy Transits and Pulsations: Empirical Variability in White Dwarfs Using Gaia and the Zwicky Transient Facility}",
      journal = {\apj},
         year = 2021,
        month = may,
       volume = {912},
       number = {2},
          eid = {125},
        pages = {125},
          doi = {10.3847/1538-4357/abee68},
archivePrefix = {arXiv},
       eprint = {2012.00035},
 primaryClass = {astro-ph.SR},
       adsurl = {https://ui.adsabs.harvard.edu/abs/2021ApJ...912..125G}
}

@ARTICLE{guretal2017,
       author = {{Gurri}, Pol and {Veras}, Dimitri and {G{\"a}nsicke}, Boris T.},
        title = "{Mass and eccentricity constraints on the planetary debris orbiting the white dwarf WD 1145+017}",
      journal = {\mnras},
         year = 2017,
        month = jan,
       volume = {464},
       number = {1},
        pages = {321-328},
          doi = {10.1093/mnras/stw2293},
archivePrefix = {arXiv},
       eprint = {1609.02563},
 primaryClass = {astro-ph.EP},
       adsurl = {https://ui.adsabs.harvard.edu/abs/2017MNRAS.464..321G}
}

@ARTICLE{ganetal2006,
       author = {{G{\"a}nsicke}, B.~T. and {Marsh}, T.~R. and {Southworth}, J. and {Rebassa-Mansergas}, A.},
        title = "{A Gaseous Metal Disk Around a White Dwarf}",
      journal = {Science},
         year = 2006,
        month = dec,
       volume = {314},
       number = {5807},
        pages = {1908},
          doi = {10.1126/science.1135033},
archivePrefix = {arXiv},
       eprint = {astro-ph/0612697},
 primaryClass = {astro-ph},
       adsurl = {https://ui.adsabs.harvard.edu/abs/2006Sci...314.1908G}
}

@ARTICLE{ganetal2007,
       author = {{G{\"a}nsicke}, B.~T. and {Marsh}, T.~R. and {Southworth}, J.},
        title = "{SDSSJ104341.53+085558.2: a second white dwarf with a gaseous debris disc}",
      journal = {\mnras},
         year = 2007,
        month = sep,
       volume = {380},
       number = {1},
        pages = {L35-L39},
          doi = {10.1111/j.1745-3933.2007.00343.x},
archivePrefix = {arXiv},
       eprint = {0705.0447},
 primaryClass = {astro-ph},
       adsurl = {https://ui.adsabs.harvard.edu/abs/2007MNRAS.380L..35G}
}

@ARTICLE{ganetal2019,
       author = {{G{\"a}nsicke}, Boris T. and {Schreiber}, Matthias R. and {Toloza}, Odette and {Gentile Fusillo}, Nicola P. and {Koester}, Detlev and {Manser}, Christopher J.},
        title = "{Accretion of a giant planet onto a white dwarf star}",
      journal = {\nat},
         year = 2019,
        month = dec,
       volume = {576},
       number = {7785},
        pages = {61-64},
          doi = {10.1038/s41586-019-1789-8},
archivePrefix = {arXiv},
       eprint = {1912.01611},
 primaryClass = {astro-ph.EP},
       adsurl = {https://ui.adsabs.harvard.edu/abs/2019Natur.576...61G}
}

@ARTICLE{grolam2003,
       author = {{Groussin}, O. and {Lamy}, P.},
        title = "{Activity on the surface of the nucleus of comet 46P/Wirtanen}",
      journal = {\aap},
         year = 2003,
        month = dec,
       volume = {412},
        pages = {879-891},
          doi = {10.1051/0004-6361:20031496},
       adsurl = {https://ui.adsabs.harvard.edu/abs/2003A&A...412..879G}
}

@ARTICLE{guietal2025,
       author = {{Guidry}, Joseph A. and {Vanderbosch}, Zachary P. and {Hermes}, J.~J. and {Veras}, Dimitri and {Hollands}, Mark A. and {Bhattacharjee}, Soumyadeep and {Caiazzo}, Ilaria and {El-Badry}, Kareem and {Kao}, Malia L. and {Ould Rouis}, Lou Baya and {Rodriguez}, Antonio C. and {van Roestel}, Jan},
        title = "{Transiting Planetary Debris near the Roche Limit of a White Dwarf on a 4.97$\,$hr Orbit -- and its Vanishing}",
      journal = {arXiv e-prints},
         year = 2025,
        month = aug,
          eid = {arXiv:2508.18348},
        pages = {arXiv:2508.18348},
          doi = {10.48550/arXiv.2508.18348},
archivePrefix = {arXiv},
       eprint = {2508.18348},
 primaryClass = {astro-ph.SR},
       adsurl = {https://ui.adsabs.harvard.edu/abs/2025arXiv250818348G}
}

@ARTICLE{gunetal2011,
       author = {{Gundlach}, B. and {Skorov}, Yu. V. and {Blum}, J.},
        title = "{Outgassing of icy bodies in the Solar System - I. The sublimation of hexagonal water ice through dust layers}",
      journal = {\icarus},
         year = 2011,
        month = jun,
       volume = {213},
       number = {2},
        pages = {710-719},
          doi = {10.1016/j.icarus.2011.03.022},
archivePrefix = {arXiv},
       eprint = {1101.2518},
 primaryClass = {astro-ph.EP},
       adsurl = {https://ui.adsabs.harvard.edu/abs/2011Icar..213..710G}
}

@ARTICLE{heretal2025,
       author = {{Hermes}, J.~J. and {Guidry}, Joseph A. and {Vanderbosch}, Zachary P. and {Badenas-Agusti}, Mariona and {Xu}, Siyi and {Kao}, Malia L. and {Rodriguez}, Antonio C. and {Hawkins}, Keith},
        title = "{Sporadic Dips from Extended Debris Transiting the Metal-rich White Dwarf SBSS 1232+563}",
      journal = {\apj},
         year = 2025,
        month = feb,
       volume = {980},
       number = {1},
          eid = {56},
        pages = {56},
          doi = {10.3847/1538-4357/ada5fd},
archivePrefix = {arXiv},
       eprint = {2501.02050},
 primaryClass = {astro-ph.SR},
       adsurl = {https://ui.adsabs.harvard.edu/abs/2025ApJ...980...56H}
}

@ARTICLE{holetal2018,
       author = {{Hollands}, M.~A. and {G{\"a}nsicke}, B.~T. and {Koester}, D.},
        title = "{Cool DZ white dwarfs II: compositions and evolution of old remnant planetary systems}",
      journal = {\mnras},
         year = 2018,
        month = jun,
       volume = {477},
       number = {1},
        pages = {93-111},
          doi = {10.1093/mnras/sty592},
archivePrefix = {arXiv},
       eprint = {1801.07714},
 primaryClass = {astro-ph.SR},
       adsurl = {https://ui.adsabs.harvard.edu/abs/2018MNRAS.477...93H}
}

@ARTICLE{hosetal2020,
       author = {{Hoskin}, Matthew J. and {Toloza}, Odette and {G{\"a}nsicke}, Boris T. and {Raddi}, Roberto and {Koester}, Detlev and {Pala}, Anna F. and {Manser}, Christopher J. and {Farihi}, Jay and {Belmonte}, Maria Teresa and {Hollands}, Mark and {Gentile Fusillo}, Nicola and {Swan}, Andrew},
        title = "{White dwarf pollution by hydrated planetary remnants: hydrogen and metals in WD J204713.76-125908.9}",
      journal = {\mnras},
         year = 2020,
        month = nov,
       volume = {499},
       number = {1},
        pages = {171-182},
          doi = {10.1093/mnras/staa2717},
archivePrefix = {arXiv},
       eprint = {2009.05053},
 primaryClass = {astro-ph.EP},
       adsurl = {https://ui.adsabs.harvard.edu/abs/2020MNRAS.499..171H}
}

@ARTICLE{jewitt2025,
       author = {{Jewitt}, David},
        title = "{Nongravitational Forces in Planetary Systems}",
      journal = {Planetary Science Journal},
         year = 2025,
        month = jan,
       volume = {6},
       number = {1},
          eid = {12},
        pages = {12},
          doi = {10.3847/PSJ/ad9824},
archivePrefix = {arXiv},
       eprint = {2411.10923},
 primaryClass = {astro-ph.EP},
       adsurl = {https://ui.adsabs.harvard.edu/abs/2025PSJ.....6...12J}
}

@ARTICLE{johetal2022,
       author = {{Johnson}, Ted M. and {Klein}, Beth L. and {Koester}, D. and {Melis}, Carl and {Zuckerman}, B. and {Jura}, M.},
        title = "{Unusual Abundances from Planetary System Material Polluting the White Dwarf G238-44}",
      journal = {\apj},
         year = 2022,
        month = dec,
       volume = {941},
       number = {2},
          eid = {113},
        pages = {113},
          doi = {10.3847/1538-4357/aca089},
archivePrefix = {arXiv},
       eprint = {2211.02673},
 primaryClass = {astro-ph.EP},
       adsurl = {https://ui.adsabs.harvard.edu/abs/2022ApJ...941..113J}
}

@ARTICLE{jura2003,
       author = {{Jura}, M.},
        title = "{A Tidally Disrupted Asteroid around the White Dwarf G29-38}",
      journal = {\apjl},
         year = 2003,
        month = feb,
       volume = {584},
       number = {2},
        pages = {L91-L94},
          doi = {10.1086/374036},
archivePrefix = {arXiv},
       eprint = {astro-ph/0301411},
 primaryClass = {astro-ph},
       adsurl = {https://ui.adsabs.harvard.edu/abs/2003ApJ...584L..91J}
}

@ARTICLE{jurxu2010,
       author = {{Jura}, M. and {Xu}, S.},
        title = "{The Survival of Water Within Extrasolar Minor Planets}",
      journal = {\aj},
         year = 2010,
        month = nov,
       volume = {140},
       number = {5},
        pages = {1129-1136},
          doi = {10.1088/0004-6256/140/5/1129},
archivePrefix = {arXiv},
       eprint = {1001.2595},
 primaryClass = {astro-ph.EP},
       adsurl = {https://ui.adsabs.harvard.edu/abs/2010AJ....140.1129J}
}

@ARTICLE{juretal2012,
       author = {{Jura}, M. and {Xu}, S. and {Klein}, B. and {Koester}, D. and {Zuckerman}, B.},
        title = "{Two Extrasolar Asteroids with Low Volatile-element Mass Fractions}",
      journal = {\apj},
         year = 2012,
        month = may,
       volume = {750},
       number = {1},
          eid = {69},
        pages = {69},
          doi = {10.1088/0004-637X/750/1/69},
archivePrefix = {arXiv},
       eprint = {1203.2885},
 primaryClass = {astro-ph.EP},
       adsurl = {https://ui.adsabs.harvard.edu/abs/2012ApJ...750...69J}
}

@ARTICLE{katz2018,
       author = {{Katz}, J.~I.},
        title = "{Why is interstellar object 1I/2017 U1 (`Oumuamua) rocky, tumbling and possibly very prolate?}",
      journal = {\mnras},
         year = 2018,
        month = jul,
       volume = {478},
       number = {1},
        pages = {L95-L98},
          doi = {10.1093/mnrasl/sly074},
archivePrefix = {arXiv},
       eprint = {1802.02273},
 primaryClass = {astro-ph.EP},
       adsurl = {https://ui.adsabs.harvard.edu/abs/2018MNRAS.478L..95K}
}

@ARTICLE{keletal2015,
       author = {{Keller}, H.~U. and {Mottola}, S. and {Davidsson}, B. and {Schr{\"o}der}, S.~E. and {Skorov}, Y. and {K{\"u}hrt}, E. and {Groussin}, O. and {Pajola}, M. and {Hviid}, S.~F. and {Preusker}, F. and {Scholten}, F. and {A'Hearn}, M.~F. and {Sierks}, H. and {Barbieri}, C. and {Lamy}, P. and {Rodrigo}, R. and {Koschny}, D. and {Rickman}, H. and {Barucci}, M.~A. and {Bertaux}, J. -L. and {Bertini}, I. and {Cremonese}, G. and {Da Deppo}, V. and {Debei}, S. and {De Cecco}, M. and {Fornasier}, S. and {Fulle}, M. and {Guti{\'e}rrez}, P.~J. and {Ip}, W. -H. and {Jorda}, L. and {Knollenberg}, J. and {Kramm}, J.~R. and {K{\"u}ppers}, M. and {Lara}, L.~M. and {Lazzarin}, M. and {Lopez Moreno}, J.~J. and {Marzari}, F. and {Michalik}, H. and {Naletto}, G. and {Sabau}, L. and {Thomas}, N. and {Vincent}, J. -B. and {Wenzel}, K. -P. and {Agarwal}, J. and {G{\"u}ttler}, C. and {Oklay}, N. and {Tubiana}, C.},
        title = "{Insolation, erosion, and morphology of comet 67P/Churyumov-Gerasimenko}",
      journal = {\aap},
         year = 2015,
        month = nov,
       volume = {583},
          eid = {A34},
        pages = {A34},
          doi = {10.1051/0004-6361/201525964},
       adsurl = {https://ui.adsabs.harvard.edu/abs/2015A&A...583A..34K}
}

@ARTICLE{kisetal2023,
       author = {{Kislyakova}, K.~G. and {Noack}, L. and {Sanchis}, E. and {Fossati}, L. and {Valyavin}, G.~G. and {Golabek}, G.~J. and {G{\"u}del}, M.},
        title = "{Induction heating of planetary interiors in white dwarf systems}",
      journal = {\aap},
         year = 2023,
        month = sep,
       volume = {677},
          eid = {A109},
        pages = {A109},
          doi = {10.1051/0004-6361/202245225},
       adsurl = {https://ui.adsabs.harvard.edu/abs/2023A&A...677A.109K}
}

@INCOLLECTION{knikok2024,
       author = {{Knight}, Matthew M. and {Kokotanekova}, Rosita and {Samarasinha}, Nalin H.},
        title = "{Physical and Surface Properties of Comet Nuclei from Remote Observations}",
    booktitle = {Comets III},
    publisher= {University of Arizona Press},
         year = 2024,
       editor = {{Meech}, Karen. J. and {Combi}, Michael. R. and {Bockel{\'e}e-Morvan}, Dominique and {Raymodn}, Sean. N. and {Zolensky}, Michael. E.},
        pages = {361-404},
          doi = {10.2458/azu_uapress_9780816553631-ch012},
       adsurl = {https://ui.adsabs.harvard.edu/abs/2024come.book..361K}
}

@ARTICLE{koester2009,
       author = {{Koester}, D.},
        title = "{Accretion and diffusion in white dwarfs. New diffusion timescales and applications to GD 362 and G 29-38}",
      journal = {\aap},
         year = 2009,
        month = may,
       volume = {498},
       number = {2},
        pages = {517-525},
          doi = {10.1051/0004-6361/200811468},
archivePrefix = {arXiv},
       eprint = {0903.1499},
 primaryClass = {astro-ph.SR},
       adsurl = {https://ui.adsabs.harvard.edu/abs/2009A&A...498..517K}
}

@ARTICLE{koeetal2014,
       author = {{Koester}, D. and {G{\"a}nsicke}, B.~T. and {Farihi}, J.},
        title = "{The frequency of planetary debris around young white dwarfs}",
      journal = {\aap},
         year = 2014,
        month = jun,
       volume = {566},
          eid = {A34},
        pages = {A34},
          doi = {10.1051/0004-6361/201423691},
archivePrefix = {arXiv},
       eprint = {1404.2617},
 primaryClass = {astro-ph.SR},
       adsurl = {https://ui.adsabs.harvard.edu/abs/2014A&A...566A..34K}
}

@ARTICLE{langmuir1913,
       author = {{Langmuir}, Irving},
        title = "{The Vapor Pressure of Metallic Tungsten}",
      journal = {Physical Review},
         year = 1913,
        month = nov,
       volume = {2},
       number = {5},
        pages = {329-342},
          doi = {10.1103/PhysRev.2.329},
       adsurl = {https://ui.adsabs.harvard.edu/abs/1913PhRv....2..329L}
}

@ARTICLE{lauetal2019,
       author = {{Lauretta}, D.~S. and {Hergenrother}, C.~W. and {Chesley}, S.~R. and {Leonard}, J.~M. and {Pelgrift}, J.~Y. and {Adam}, C.~D. and {Al Asad}, M. and {Antreasian}, P.~G. and {Ballouz}, R. -L. and {Becker}, K.~J. and {Bennett}, C.~A. and {Bos}, B.~J. and {Bottke}, W.~F. and {Brozovi{\'c}}, M. and {Campins}, H. and {Connolly}, H.~C. and {Daly}, M.~G. and {Davis}, A.~B. and {de Le{\'o}n}, J. and {DellaGiustina}, D.~N. and {Drouet d'Aubigny}, C.~Y. and {Dworkin}, J.~P. and {Emery}, J.~P. and {Farnocchia}, D. and {Glavin}, D.~P. and {Golish}, D.~R. and {Hartzell}, C.~M. and {Jacobson}, R.~A. and {Jawin}, E.~R. and {Jenniskens}, P. and {Kidd}, J.~N. and {Lessac-Chenen}, E.~J. and {Li}, J. -Y. and {Libourel}, G. and {Licandro}, J. and {Liounis}, A.~J. and {Maleszewski}, C.~K. and {Manzoni}, C. and {May}, B. and {McCarthy}, L.~K. and {McMahon}, J.~W. and {Michel}, P. and {Molaro}, J.~L. and {Moreau}, M.~C. and {Nelson}, D.~S. and {Owen}, W.~M. and {Rizk}, B. and {Roper}, H.~L. and {Rozitis}, B. and {Sahr}, E.~M. and {Scheeres}, D.~J. and {Seabrook}, J.~A. and {Selznick}, S.~H. and {Takahashi}, Y. and {Thuillet}, F. and {Tricarico}, P. and {Vokrouhlick{\'y}}, D. and {Wolner}, C.~W.~V.},
        title = "{Episodes of particle ejection from the surface of the active asteroid (101955) Bennu}",
      journal = {Science},
         year = 2019,
        month = dec,
       volume = {366},
       number = {6470},
          eid = {eaay3544},
        pages = {eaay3544},
          doi = {10.1126/science.aay3544},
       adsurl = {https://ui.adsabs.harvard.edu/abs/2019Sci...366.3544L}
}

@ARTICLE{yuqilietal2024,
       author = {{Li}, Yuqi and {Bonsor}, Amy and {Shorttle}, Oliver},
        title = "{Post-main sequence thermal evolution of planetesimals}",
      journal = {\mnras},
         year = 2024,
        month = jan,
       volume = {527},
       number = {1},
        pages = {1014-1032},
          doi = {10.1093/mnras/stad3131},
archivePrefix = {arXiv},
       eprint = {2310.17057},
 primaryClass = {astro-ph.EP},
       adsurl = {https://ui.adsabs.harvard.edu/abs/2024MNRAS.527.1014L}
}

@ARTICLE{lietal2025a,
       author = {{Li}, Yuqi and {Bonsor}, Amy and {Shorttle}, Oliver and {Rogers}, Laura K.},
        title = "{Can tidal evolution lead to close-in planetary bodies around white dwarfs - I. Orbital period distribution}",
      journal = {\mnras},
         year = 2025,
        month = feb,
       volume = {537},
       number = {2},
        pages = {2214-2231},
          doi = {10.1093/mnras/staf182},
archivePrefix = {arXiv},
       eprint = {2506.20301},
 primaryClass = {astro-ph.EP},
       adsurl = {https://ui.adsabs.harvard.edu/abs/2025MNRAS.537.2214L}
}

@ARTICLE{lietal2025b,
       author = {{Li}, Yuqi and {Bonsor}, Amy and {Shorttle}, Oliver},
        title = "{Can tidal evolution lead to close-in planetary bodies around white dwarfs {\textendash} II. Volcanism and transits}",
      journal = {\mnras},
         year = 2025,
        month = jul,
       volume = {541},
       number = {1},
        pages = {610-629},
          doi = {10.1093/mnras/staf1028},
archivePrefix = {arXiv},
       eprint = {2506.20316},
 primaryClass = {astro-ph.EP},
       adsurl = {https://ui.adsabs.harvard.edu/abs/2025MNRAS.541..610L}
}

@ARTICLE{limetal2024,
       author = {{Limbach}, Mary Anne and {Vanderburg}, Andrew and {Venner}, Alexander and {Blouin}, Simon and {Stevenson}, Kevin B. and {MacDonald}, Ryan J. and {Jenkins}, Sydney and {Bowens-Rubin}, Rachel and {Soares-Furtado}, Melinda and {Morley}, Caroline and {Janson}, Markus and {Debes}, John and {Xu}, Siyi and {Kleisioti}, Evangelia and {Kenworthy}, Matthew and {Butler}, Paul and {Crane}, Jeffrey D. and {Osip}, Dave and {Shectman}, Stephen and {Teske}, Johanna},
        title = "{The MIRI Exoplanets Orbiting White dwarfs (MEOW) Survey: Mid-infrared Excess Reveals a Giant Planet Candidate around a Nearby White Dwarf}",
      journal = {\apjl},
         year = 2024,
        month = sep,
       volume = {973},
       number = {1},
          eid = {L11},
        pages = {L11},
          doi = {10.3847/2041-8213/ad74ed},
archivePrefix = {arXiv},
       eprint = {2408.16813},
 primaryClass = {astro-ph.EP},
       adsurl = {https://ui.adsabs.harvard.edu/abs/2024ApJ...973L..11L}
}

@ARTICLE{luhetal2011,
       author = {{Luhman}, K.~L. and {Burgasser}, A.~J. and {Bochanski}, J.~J.},
        title = "{Discovery of a Candidate for the Coolest Known Brown Dwarf}",
      journal = {\apjl},
         year = 2011,
        month = mar,
       volume = {730},
       number = {1},
          eid = {L9},
        pages = {L9},
          doi = {10.1088/2041-8205/730/1/L9},
archivePrefix = {arXiv},
       eprint = {1102.5411},
 primaryClass = {astro-ph.GA},
       adsurl = {https://ui.adsabs.harvard.edu/abs/2011ApJ...730L...9L}
}

@ARTICLE{malper2016,
       author = {{Malamud}, Uri and {Perets}, Hagai B.},
        title = "{Post-main Sequence Evolution of Icy Minor Planets: Implications for Water Retention and White Dwarf Pollution}",
      journal = {\apj},
         year = 2016,
        month = dec,
       volume = {832},
       number = {2},
          eid = {160},
        pages = {160},
          doi = {10.3847/0004-637X/832/2/160},
archivePrefix = {arXiv},
       eprint = {1608.00593},
 primaryClass = {astro-ph.SR},
       adsurl = {https://ui.adsabs.harvard.edu/abs/2016ApJ...832..160M}
}

@ARTICLE{malper2017a,
       author = {{Malamud}, Uri and {Perets}, Hagai B.},
        title = "{Post-main-sequence Evolution of Icy Minor Planets. II. Water Retention and White Dwarf Pollution around Massive Progenitor Stars}",
      journal = {\apj},
         year = 2017,
        month = jun,
       volume = {842},
       number = {1},
          eid = {67},
        pages = {67},
          doi = {10.3847/1538-4357/aa7055},
archivePrefix = {arXiv},
       eprint = {1704.01165},
 primaryClass = {astro-ph.EP},
       adsurl = {https://ui.adsabs.harvard.edu/abs/2017ApJ...842...67M}
}

@ARTICLE{malper2017b,
       author = {{Malamud}, Uri and {Perets}, Hagai B.},
        title = "{Post-main-sequence Evolution of Icy Minor Planets. III. Water Retention in Dwarf Planets and Exomoons and Implications for White Dwarf Pollution}",
      journal = {\apj},
         year = 2017,
        month = nov,
       volume = {849},
       number = {1},
          eid = {8},
        pages = {8},
          doi = {10.3847/1538-4357/aa8df5},
archivePrefix = {arXiv},
       eprint = {1708.07489},
 primaryClass = {astro-ph.SR},
       adsurl = {https://ui.adsabs.harvard.edu/abs/2017ApJ...849....8M}
}

@ARTICLE{malper2020a,
       author = {{Malamud}, Uri and {Perets}, Hagai B.},
        title = "{Tidal disruption of planetary bodies by white dwarfs I: a hybrid SPH-analytical approach}",
      journal = {\mnras},
         year = 2020,
        month = mar,
       volume = {492},
       number = {4},
        pages = {5561-5581},
          doi = {10.1093/mnras/staa142},
archivePrefix = {arXiv},
       eprint = {1911.12068},
 primaryClass = {astro-ph.EP},
       adsurl = {https://ui.adsabs.harvard.edu/abs/2020MNRAS.492.5561M}
}

@ARTICLE{malper2020b,
       author = {{Malamud}, Uri and {Perets}, Hagai B.},
        title = "{Tidal disruption of planetary bodies by white dwarfs - II. Debris disc structure and ejected interstellar asteroids}",
      journal = {\mnras},
         year = 2020,
        month = mar,
       volume = {493},
       number = {1},
        pages = {698-712},
          doi = {10.1093/mnras/staa143},
archivePrefix = {arXiv},
       eprint = {1911.12184},
 primaryClass = {astro-ph.EP},
       adsurl = {https://ui.adsabs.harvard.edu/abs/2020MNRAS.493..698M}
}

@INPROCEEDINGS{malamud2026,
       author = {{Malamud}, Uri},
        title = "{White dwarf systems: Exoplanets and debris disks}",
    booktitle = {Encyclopedia of Astrophysics},
         year = 2026,
       volume = {1},
        month = jan,
        pages = {553-563},
          doi = {10.1016/B978-0-443-21439-4.00001-8},
       adsurl = {https://ui.adsabs.harvard.edu/abs/2026enap....1..553M}
}

@ARTICLE{manetal2019,
       author = {{Manser}, Christopher J. and {G{\"a}nsicke}, Boris T. and {Eggl}, Siegfried and {Hollands}, Mark and {Izquierdo}, Paula and {Koester}, Detlev and {Landstreet}, John D. and {Lyra}, Wladimir and {Marsh}, Thomas R. and {Meru}, Farzana and {Mustill}, Alexander J. and {Rodr{\'\i}guez-Gil}, Pablo and {Toloza}, Odette and {Veras}, Dimitri and {Wilson}, David J. and {Burleigh}, Matthew R. and {Davies}, Melvyn B. and {Farihi}, Jay and {Gentile Fusillo}, Nicola and {de Martino}, Domitilla and {Parsons}, Steven G. and {Quirrenbach}, Andreas and {Raddi}, Roberto and {Reffert}, Sabine and {Del Santo}, Melania and {Schreiber}, Matthias R. and {Silvotti}, Roberto and {Toonen}, Silvia and {Villaver}, Eva and {Wyatt}, Mark and {Xu}, Siyi and {Portegies Zwart}, Simon},
        title = "{A planetesimal orbiting within the debris disc around a white dwarf star}",
      journal = {Science},
         year = 2019,
        month = apr,
       volume = {364},
       number = {6435},
        pages = {66-69},
          doi = {10.1126/science.aat5330},
archivePrefix = {arXiv},
       eprint = {1904.02163},
 primaryClass = {astro-ph.EP},
       adsurl = {https://ui.adsabs.harvard.edu/abs/2019Sci...364...66M}
}

@ARTICLE{manetal2020,
       author = {{Manser}, Christopher J. and {G{\"a}nsicke}, Boris T. and {Gentile Fusillo}, Nicola Pietro and {Ashley}, Richard and {Breedt}, Elm{\'e} and {Hollands}, Mark and {Izquierdo}, Paula and {Pelisoli}, Ingrid},
        title = "{The frequency of gaseous debris discs around white dwarfs}",
      journal = {\mnras},
         year = 2020,
        month = apr,
       volume = {493},
       number = {2},
        pages = {2127-2139},
          doi = {10.1093/mnras/staa359},
archivePrefix = {arXiv},
       eprint = {2002.01936},
 primaryClass = {astro-ph.EP},
       adsurl = {https://ui.adsabs.harvard.edu/abs/2020MNRAS.493.2127M}
}

@ARTICLE{mcdver2021,
       author = {{McDonald}, Catriona H. and {Veras}, Dimitri},
        title = "{White dwarf planetary debris dependence on physical structure distributions within asteroid belts}",
      journal = {\mnras},
         year = 2021,
        month = sep,
       volume = {506},
       number = {3},
        pages = {4031-4047},
          doi = {10.1093/mnras/stab1906},
archivePrefix = {arXiv},
       eprint = {2107.00322},
 primaryClass = {astro-ph.EP},
       adsurl = {https://ui.adsabs.harvard.edu/abs/2021MNRAS.506.4031M}
}

@ARTICLE{meletal2020,
       author = {{Melis}, Carl and {Klein}, Beth and {Doyle}, Alexandra E. and {Weinberger}, Alycia and {Zuckerman}, B. and {Dufour}, Patrick},
        title = "{Serendipitous Discovery of Nine White Dwarfs with Gaseous Debris Disks}",
      journal = {\apj},
         year = 2020,
        month = dec,
       volume = {905},
       number = {1},
          eid = {56},
        pages = {56},
          doi = {10.3847/1538-4357/abbdfa},
archivePrefix = {arXiv},
       eprint = {2010.03695},
 primaryClass = {astro-ph.SR},
       adsurl = {https://ui.adsabs.harvard.edu/abs/2020ApJ...905...56M}
}

@ARTICLE{mestel1952,
       author = {{Mestel}, L.},
        title = "{On the theory of white dwarf stars. I. The energy sources of white dwarfs}",
      journal = {\mnras},
         year = 1952,
        month = jan,
       volume = {112},
        pages = {583},
          doi = {10.1093/mnras/112.6.583},
       adsurl = {https://ui.adsabs.harvard.edu/abs/1952MNRAS.112..583M}
}

@ARTICLE{moletal2020,
       author = {{Molaro}, J.~L. and {Hergenrother}, C.~W. and {Chesley}, S.~R. and {Walsh}, K.~J. and {Hanna}, R.~D. and {Haberle}, C.~W. and {Schwartz}, S.~R. and {Ballouz}, R. -L. and {Bottke}, W.~F. and {Campins}, H.~J. and {Lauretta}, D.~S.},
        title = "{Thermal Fatigue as a Driving Mechanism for Activity on Asteroid Bennu}",
      journal = {Journal of Geophysical Research (Planets)},
         year = 2020,
        month = aug,
       volume = {125},
       number = {8},
          eid = {e06325},
        pages = {e06325},
          doi = {10.1029/2019JE00632510.1002/essoar.10501385.2},
       adsurl = {https://ui.adsabs.harvard.edu/abs/2020JGRE..12506325M}
}

@ARTICLE{muesam2018,
       author = {{Mueller}, Beatrice E.~A. and {Samarasinha}, Nalin H.},
        title = "{Further Investigation of Changes in Cometary Rotation}",
      journal = {\aj},
         year = 2018,
        month = sep,
       volume = {156},
       number = {3},
          eid = {107},
        pages = {107},
          doi = {10.3847/1538-3881/aad0a1},
archivePrefix = {arXiv},
       eprint = {1806.11158},
 primaryClass = {astro-ph.EP},
       adsurl = {https://ui.adsabs.harvard.edu/abs/2018AJ....156..107M}
}

@ARTICLE{nauenberg1972,
       author = {{Nauenberg}, Michael},
        title = "{Analytic Approximations to the Mass-Radius Relation and Energy of Zero-Temperature Stars}",
      journal = {\apj},
         year = 1972,
        month = jul,
       volume = {175},
        pages = {417},
          doi = {10.1086/151568},
       adsurl = {https://ui.adsabs.harvard.edu/abs/1972ApJ...175..417N}
}

@ARTICLE{nooetal2025,
       author = {{Noor}, Hiba Tu and {Farihi}, Jay and {Kenyon}, Scott J. and {Rafikov}, Roman R. and {Wyatt}, Mark C. and {Su}, Kate Y.~L. and {Melis}, Carl and {Swan}, Andrew and {Wilson}, Thomas G. and {G{\"a}nsicke}, Boris T. and {Bonsor}, Amy and {Rogers}, Laura K. and {Redfield}, Seth and {Kilic}, Mukremin},
        title = "{Activity in White Dwarf Debris Disks I: Spitzer Legacy Reveals Variability Incompatible with the Canonical Model}",
      journal = {\mnras},
         year = 2025,
        month = aug,
          doi = {10.1093/mnras/staf1380},
archivePrefix = {arXiv},
       eprint = {2508.13119},
 primaryClass = {astro-ph.EP},
       adsurl = {https://ui.adsabs.harvard.edu/abs/2025MNRAS.tmp.1320N}
}

@ARTICLE{obretal2024,
       author = {{O'Brien}, Mairi W. and {Tremblay}, P. -E. and {Klein}, B.~L. and {Koester}, D. and {Melis}, C. and {B{\'e}dard}, A. and {Cukanovaite}, E. and {Cunningham}, T. and {Doyle}, A.~E. and {G{\"a}nsicke}, B.~T. and {Gentile Fusillo}, N.~P. and {Hollands}, M.~A. and {McCleery}, J. and {Pelisoli}, I. and {Toonen}, S. and {Weinberger}, A.~J. and {Zuckerman}, B.},
        title = "{The 40 pc sample of white dwarfs from Gaia}",
      journal = {\mnras},
         year = 2024,
        month = jan,
       volume = {527},
       number = {3},
        pages = {8687-8705},
          doi = {10.1093/mnras/stad3773},
archivePrefix = {arXiv},
       eprint = {2312.02735},
 primaryClass = {astro-ph.SR},
       adsurl = {https://ui.adsabs.harvard.edu/abs/2024MNRAS.527.8687O}
}

@ARTICLE{okuetal2023,
       author = {{Okuya}, Ayaka and {Ida}, Shigeru and {Hyodo}, Ryuki and {Okuzumi}, Satoshi},
        title = "{Modelling the evolution of silicate/volatile accretion discs around white dwarfs}",
      journal = {\mnras},
         year = 2023,
        month = feb,
       volume = {519},
       number = {2},
        pages = {1657-1676},
          doi = {10.1093/mnras/stac3522},
archivePrefix = {arXiv},
       eprint = {2211.16797},
 primaryClass = {astro-ph.EP},
       adsurl = {https://ui.adsabs.harvard.edu/abs/2023MNRAS.519.1657O}
}

@ARTICLE{ouletal2024,
       author = {{Ould Rouis}, Lou Baya and {Hermes}, J.~J. and {G{\"a}nsicke}, Boris T. and {Sahu}, Snehalata and {Koester}, Detlev and {Tremblay}, P.-E. and {Veras}, Dimitri and {Farihi}, Jay and {Heintz}, Tyler M. and {Gentile Fusillo}, Nicola Pietro and {Redfield}, Seth},
        title = "{Constraints on Remnant Planetary Systems as a Function of Main-sequence Mass with HST/COS}",
      journal = {\apj},
         year = 2024,
        month = dec,
       volume = {976},
       number = {2},
          eid = {156},
        pages = {156},
          doi = {10.3847/1538-4357/ad86bb},
archivePrefix = {arXiv},
       eprint = {2410.06335},
 primaryClass = {astro-ph.SR},
       adsurl = {https://ui.adsabs.harvard.edu/abs/2024ApJ...976..156O}
}

@ARTICLE{oweetal2023,
       author = {{Owens}, Dylan and {Xu}, Siyi and {Manjavacas}, Elena and {Leggett}, S.~K. and {Casewell}, S.~L. and {Dennihy}, Erik and {Dufour}, Patrick and {Klein}, Beth L. and {Yeh}, Sherry and {Zuckerman}, B.},
        title = "{Disk or Companion: Characterizing Excess Infrared Flux in Seven White Dwarf Systems with Near-infrared Spectroscopy}",
      journal = {\aj},
         year = 2023,
        month = jul,
       volume = {166},
       number = {1},
          eid = {5},
        pages = {5},
          doi = {10.3847/1538-3881/accc25},
archivePrefix = {arXiv},
       eprint = {2303.16330},
 primaryClass = {astro-ph.SR},
       adsurl = {https://ui.adsabs.harvard.edu/abs/2023AJ....166....5O}
}

@ARTICLE{prahar2000,
       author = {{Pravec}, Petr and {Harris}, Alan W.},
        title = "{Fast and Slow Rotation of Asteroids}",
      journal = {\icarus},
         year = 2000,
        month = nov,
       volume = {148},
       number = {1},
        pages = {12-20},
          doi = {10.1006/icar.2000.6482},
       adsurl = {https://ui.adsabs.harvard.edu/abs/2000Icar..148...12P}
}

@ARTICLE{putxu2021,
       author = {{Putirka}, Keith D. and {Xu}, Siyi},
        title = "{Polluted white dwarfs reveal exotic mantle rock types on exoplanets in our solar neighborhood}",
      journal = {Nature Communications},
         year = 2021,
        month = nov,
       volume = {12},
          eid = {6168},
        pages = {6168},
          doi = {10.1038/s41467-021-26403-8},
archivePrefix = {arXiv},
       eprint = {2111.03124},
 primaryClass = {astro-ph.EP},
       adsurl = {https://ui.adsabs.harvard.edu/abs/2021NatCo..12.6168P}
}

@ARTICLE{radetal2015,
       author = {{Raddi}, R. and {G{\"a}nsicke}, B.~T. and {Koester}, D. and {Farihi}, J. and {Hermes}, J.~J. and {Scaringi}, S. and {Breedt}, E. and {Girven}, J.},
        title = "{Likely detection of water-rich asteroid debris in a metal-polluted white dwarf}",
      journal = {\mnras},
         year = 2015,
        month = jun,
       volume = {450},
       number = {2},
        pages = {2083-2093},
          doi = {10.1093/mnras/stv701},
archivePrefix = {arXiv},
       eprint = {1503.07864},
 primaryClass = {astro-ph.SR},
       adsurl = {https://ui.adsabs.harvard.edu/abs/2015MNRAS.450.2083R}
}

@ARTICLE{rafikov2011a,
       author = {{Rafikov}, Roman R.},
        title = "{Metal Accretion onto White Dwarfs Caused by Poynting-Robertson Drag on their Debris Disks}",
      journal = {\apjl},
         year = 2011,
        month = may,
       volume = {732},
       number = {1},
          eid = {L3},
        pages = {L3},
          doi = {10.1088/2041-8205/732/1/L3},
archivePrefix = {arXiv},
       eprint = {1102.3153},
 primaryClass = {astro-ph.EP},
       adsurl = {https://ui.adsabs.harvard.edu/abs/2011ApJ...732L...3R}
}

@ARTICLE{rapetal2016,
       author = {{Rappaport}, S. and {Gary}, B.~L. and {Kaye}, T. and {Vanderburg}, A. and {Croll}, B. and {Benni}, P. and {Foote}, J.},
        title = "{Drifting asteroid fragments around WD 1145+017}",
      journal = {\mnras},
         year = 2016,
        month = jun,
       volume = {458},
       number = {4},
        pages = {3904-3917},
          doi = {10.1093/mnras/stw612},
archivePrefix = {arXiv},
       eprint = {1602.00740},
 primaryClass = {astro-ph.EP},
       adsurl = {https://ui.adsabs.harvard.edu/abs/2016MNRAS.458.3904R}
}

@ARTICLE{rogetal2025,
       author = {{Rogers}, Laura K. and {Manser}, Christopher J. and {Bonsor}, Amy and {Dennihy}, Erik and {Hodgkin}, Simon and {Kissler-Patig}, Markus and {Lai}, Samuel and {Melis}, Carl and {Xu}, Siyi and {Gentile Fusillo}, Nicola and {G{\"a}nsicke}, Boris and {Swan}, Andrew and {Toloza}, Odette and {Veras}, Dimitri},
        title = "{Simultaneous emission from dust and gas in the planetary debris orbiting a white dwarf}",
      journal = {\mnras},
         year = 2025,
        month = feb,
       volume = {537},
       number = {1},
        pages = {L72-L79},
          doi = {10.1093/mnrasl/slae117},
archivePrefix = {arXiv},
       eprint = {2412.07647},
 primaryClass = {astro-ph.EP},
       adsurl = {https://ui.adsabs.harvard.edu/abs/2025MNRAS.537L..72R}
}

@ARTICLE{rubincam2000,
       author = {{Rubincam}, David Parry},
        title = "{Radiative Spin-up and Spin-down of Small Asteroids}",
      journal = {\icarus},
         year = 2000,
        month = nov,
       volume = {148},
       number = {1},
        pages = {2-11},
          doi = {10.1006/icar.2000.6485},
       adsurl = {https://ui.adsabs.harvard.edu/abs/2000Icar..148....2R}
}

@ARTICLE{safetal2021,
       author = {{Safrit}, Taylor K. and {Steckloff}, Jordan K. and {Bosh}, Amanda S. and {Nesvorny}, David and {Walsh}, Kevin and {Brasser}, Ramon and {Minton}, David A.},
        title = "{The Formation of Bilobate Comet Shapes through Sublimative Torques}",
      journal = {PSJ},
         year = 2021,
        month = feb,
       volume = {2},
       number = {1},
          eid = {14},
        pages = {14},
          doi = {10.3847/PSJ/abc9c8},
archivePrefix = {arXiv},
       eprint = {2011.01394},
 primaryClass = {astro-ph.EP},
       adsurl = {https://ui.adsabs.harvard.edu/abs/2021PSJ.....2...14S}
}

@ARTICLE{sahetal2025,
       author = {{Sahu}, Snehalata and {G{\"a}nsicke}, Boris T. and {Williams}, Jamie T. and {Koester}, Detlev G. and {Farihi}, Jay and {Desch}, Steven J. and {Gentile Fusillo}, Nicola Pietro and {Veras}, Dimitri and {Raymond}, Sean N. and {Belmonte}, Maria Teresa},
        title = "{Discovery of an icy and nitrogen-rich extrasolar planetesimal}",
      journal = {\mnras},
         year = 2025,
        month = oct,
       volume = {543},
       number = {1},
        pages = {223-232},
          doi = {10.1093/mnras/staf1424},
archivePrefix = {arXiv},
       eprint = {2509.13422},
 primaryClass = {astro-ph.EP},
       adsurl = {https://ui.adsabs.harvard.edu/abs/2025MNRAS.543..223S}
}

@ARTICLE{sammue2013,
       author = {{Samarasinha}, Nalin H. and {Mueller}, B{\'e}atrice E.~A.},
        title = "{Relating Changes in Cometary Rotation to Activity: Current Status and Applications to Comet C/2012 S1 (ISON)}",
      journal = {\apjl},
         year = 2013,
        month = sep,
       volume = {775},
       number = {1},
          eid = {L10},
        pages = {L10},
          doi = {10.1088/2041-8205/775/1/L10},
archivePrefix = {arXiv},
       eprint = {1309.0586},
 primaryClass = {astro-ph.EP},
       adsurl = {https://ui.adsabs.harvard.edu/abs/2013ApJ...775L..10S}
}

@ARTICLE{sansch2014,
       author = {{S{\'a}nchez}, P. and {Scheeres}, D.~J.},
        title = "{The strength of regolith and rubble pile asteroids}",
      journal = {Meteoritics \& Planetary Science},
         year = 2014,
        month = may,
       volume = {49},
       number = {5},
        pages = {788-811},
          doi = {0.48550/arXiv.1306.1622},
archivePrefix = {arXiv},
       eprint = {1306.1622},
 primaryClass = {astro-ph.EP},
       adsurl = {https://ui.adsabs.harvard.edu/abs/2014M&PS...49..788S}
}

@ARTICLE{scheeres2007,
       author = {{Scheeres}, D.~J.},
        title = "{The dynamical evolution of uniformly rotating asteroids subject to YORP}",
      journal = {\icarus},
         year = 2007,
        month = jun,
       volume = {188},
       number = {2},
        pages = {430-450},
          doi = {10.1016/j.icarus.2006.12.015},
       adsurl = {https://ui.adsabs.harvard.edu/abs/2007Icar..188..430S}
}

@ARTICLE{scheeres2018,
       author = {{Scheeres}, D.~J.},
        title = "{Disaggregation of small, cohesive rubble pile asteroids due to YORP}",
      journal = {\icarus},
         year = 2018,
        month = apr,
       volume = {304},
        pages = {183-191},
          doi = {10.1016/j.icarus.2017.05.029},
archivePrefix = {arXiv},
       eprint = {1706.03385},
 primaryClass = {astro-ph.EP},
       adsurl = {https://ui.adsabs.harvard.edu/abs/2018Icar..304..183S}
}

@ARTICLE{sheser2023,
       author = {{Shestakova}, Lyubov I. and {Serebryanskiy}, Aleksander V.},
        title = "{On the new mechanism of planetary long-period debris formation around white dwarfs}",
      journal = {\mnras},
         year = 2023,
        month = sep,
       volume = {524},
       number = {3},
        pages = {4506-4520},
          doi = {10.1093/mnras/stad2006},
       adsurl = {https://ui.adsabs.harvard.edu/abs/2023MNRAS.524.4506S}
}

@ARTICLE{sigetal2003,
       author = {{Sigurdsson}, Steinn and {Richer}, Harvey B. and {Hansen}, Brad M. and {Stairs}, Ingrid H. and {Thorsett}, Stephen E.},
        title = "{A Young White Dwarf Companion to Pulsar B1620-26: Evidence for Early Planet Formation}",
      journal = {Science},
         year = 2003,
        month = jul,
       volume = {301},
       number = {5630},
        pages = {193-196},
          doi = {10.1126/science.1086326},
archivePrefix = {arXiv},
       eprint = {astro-ph/0307339},
 primaryClass = {astro-ph},
       adsurl = {https://ui.adsabs.harvard.edu/abs/2003Sci...301..193S}
}

@ARTICLE{steetal2015,
       author = {{Steckloff}, Jordan K. and {Johnson}, Brandon C. and {Bowling}, Timothy and {Jay Melosh}, H. and {Minton}, David and {Lisse}, Carey M. and {Battams}, Karl},
        title = "{Dynamic sublimation pressure and the catastrophic breakup of Comet ISON}",
      journal = {\icarus},
         year = 2015,
        month = sep,
       volume = {258},
        pages = {430-437},
          doi = {10.1016/j.icarus.2015.06.032},
archivePrefix = {arXiv},
       eprint = {1507.02669},
 primaryClass = {astro-ph.EP},
       adsurl = {https://ui.adsabs.harvard.edu/abs/2015Icar..258..430S}
}

@ARTICLE{stejac2016,
       author = {{Steckloff}, Jordan K. and {Jacobson}, Seth A.},
        title = "{The formation of striae within cometary dust tails by a sublimation-driven YORP-like effect}",
      journal = {\icarus},
         year = 2016,
        month = jan,
       volume = {264},
        pages = {160-171},
          doi = {10.1016/j.icarus.2015.09.021},
archivePrefix = {arXiv},
       eprint = {1509.04756},
 primaryClass = {astro-ph.EP},
       adsurl = {https://ui.adsabs.harvard.edu/abs/2016Icar..264..160S}
}

@ARTICLE{stesam2018,
       author = {{Steckloff}, Jordan K. and {Samarasinha}, Nalin H.},
        title = "{The sublimative torques of Jupiter Family Comets and mass wasting events on their nuclei}",
      journal = {\icarus},
         year = 2018,
        month = sep,
       volume = {312},
        pages = {172-180},
          doi = {10.1016/j.icarus.2018.04.031},
archivePrefix = {arXiv},
       eprint = {1804.10232},
 primaryClass = {astro-ph.EP},
       adsurl = {https://ui.adsabs.harvard.edu/abs/2018Icar..312..172S}
}

@ARTICLE{steckloffetal2021a,
       author = {{Steckloff}, Jordan K. and {Debes}, John and {Steele}, Amy and {Johnson}, Brandon and {Adams}, Elisabeth R. and {Jacobson}, Seth A. and {Springmann}, Alessondra},
        title = "{How Sublimation Delays the Onset of Dusty Debris Disk Formation around White Dwarf Stars}",
      journal = {\apjl},
         year = 2021,
        month = jun,
       volume = {913},
       number = {2},
          eid = {L31},
        pages = {L31},
          doi = {10.3847/2041-8213/abfd39},
archivePrefix = {arXiv},
       eprint = {2104.14035},
 primaryClass = {astro-ph.EP},
       adsurl = {https://ui.adsabs.harvard.edu/abs/2021ApJ...913L..31S}
}

@ARTICLE{steckloffetal2021b,
       author = {{Steckloff}, Jordan K. and {Lisse}, Carey M. and {Safrit}, Taylor K. and {Bosh}, Amanda S. and {Lyra}, Wladimir and {Sarid}, Gal},
        title = "{The sublimative evolution of (486958) Arrokoth}",
      journal = {\icarus},
         year = 2021,
        month = mar,
       volume = {356},
          eid = {113998},
        pages = {113998},
          doi = {10.1016/j.icarus.2020.113998},
archivePrefix = {arXiv},
       eprint = {2007.12657},
 primaryClass = {astro-ph.EP},
       adsurl = {https://ui.adsabs.harvard.edu/abs/2021Icar..35613998S}
}

@ARTICLE{thoetal1993,
       author = {{Thorsett}, S.~E. and {Arzoumanian}, Z. and {Taylor}, J.~H.},
        title = "{PSR B1620-26: A Binary Radio Pulsar with a Planetary Companion?}",
      journal = {\apjl},
         year = 1993,
        month = jul,
       volume = {412},
        pages = {L33},
          doi = {10.1086/186933},
       adsurl = {https://ui.adsabs.harvard.edu/abs/1993ApJ...412L..33T}
}

@ARTICLE{toretal2026,
       author = {{Torii}, Naoya and {Ida}, Shigeru and {Hyodo}, Ryuki},
        title = "{Ring formation around giant planets by tidal disruption of a single passing large Kuiper belt object II: The dynamical fate of tidal fragments}",
      journal = {arXiv e-prints},
         year = 2026,
        month = apr,
          eid = {arXiv:2604.10042},
        pages = {arXiv:2604.10042},
          doi = {10.48550/arXiv.2604.10042},
archivePrefix = {arXiv},
       eprint = {2604.10042},
 primaryClass = {astro-ph.EP},
       adsurl = {https://ui.adsabs.harvard.edu/abs/2026arXiv260410042T}
}

@ARTICLE{treetal2016,
       author = {{Tremblay}, P. -E. and {Cummings}, J. and {Kalirai}, J.~S. and {G{\"a}nsicke}, B.~T. and {Gentile-Fusillo}, N. and {Raddi}, R.},
        title = "{The field white dwarf mass distribution}",
      journal = {\mnras},
         year = 2016,
        month = sep,
       volume = {461},
       number = {2},
        pages = {2100-2114},
          doi = {10.1093/mnras/stw1447},
archivePrefix = {arXiv},
       eprint = {1606.05292},
 primaryClass = {astro-ph.SR},
       adsurl = {https://ui.adsabs.harvard.edu/abs/2016MNRAS.461.2100T}
}

%%%%%%% Don't change these lines
%%%%%%%%%%%%%%%%%%%%%%%%%%%%
\bsp
\label{lastpage}
\end{document}